\documentclass[onecolumn,journal,draftclsnofoot,12pt]{IEEEtran}

\usepackage{bm}
\usepackage{amssymb}
\usepackage{amsmath}
\usepackage{amsfonts}
\usepackage{graphicx}
\usepackage{subfigure}
\usepackage{cite}
\usepackage{enumerate}
\usepackage{url}
\usepackage{epstopdf}
\usepackage{float}
\usepackage{color}
\usepackage{mathtools}
\usepackage{multirow}
\usepackage{enumitem}
\usepackage[final]{pdfcomment}
\usepackage{enumerate}
\usepackage[rgb]{xcolor}
\usepackage{enumitem}

\usepackage[normalem]{ulem}  %下划线
\newcommand{\beq}{\begin{equation}}
\newcommand{\eeq}{\end{equation}}

\newcommand{\parhead}{\vspace{1.5mm}\noindent\textbf}

\allowdisplaybreaks[4]
\makeatletter  
\newif\if@restonecol  
\makeatother

\usepackage{amsmath}

\usepackage[linesnumbered, lined, commentsnumbered, ruled]{algorithm2e}

\usepackage{makecell}

\newif\ifshowannote
\showannotetrue

\makeatletter
\newcommand{\manuallabel}[2]{\def\@currentlabel{#2}\label{#1}}
\makeatother

\usepackage[numbers,sort&compress]{natbib}
\usepackage{etoolbox}
\makeatletter
\patchcmd{\NAT@citexnum}
  {%
    \ifx\NAT@last@yr\relax
      \def@NAT@last@yr{\@citea}%
    \else
      \def@NAT@last@yr{--\NAT@penalty}%
    \fi
  }
  {%
    \def@NAT@last@yr{--\NAT@penalty}%
  }
  {}{\FAIL}

\newcommand{\obj}{\mathrm{obj}}
\newcommand{\ris}{metasurface}
\newcommand{\Ris}{Metasurface}
\newcommand{\holosketch}{MetaSketch}

\begin{document}

\title{MetaSketch: Wireless Semantic Segmentation by Metamaterial Surfaces}

\author{
\IEEEauthorblockN{
\normalsize{Jingzhi~Hu},~\IEEEmembership{\normalsize Student~Member,~IEEE},
\normalsize{Hongliang~Zhang},~\IEEEmembership{\normalsize Member,~IEEE},
\normalsize{Kaigui~Bian},~\IEEEmembership{\normalsize Senior~Member,~IEEE},
\normalsize{Zhu~Han},~\IEEEmembership{\normalsize Fellow,~IEEE},
\normalsize{H.~Vincent~Poor},~\IEEEmembership{\normalsize Life~Fellow,~IEEE},
and~\normalsize{Lingyang~Song},~\IEEEmembership{\normalsize Fellow,~IEEE}
}
\thanks{
This work has been submitted to the IEEE for possible publication.
Copyright may be transferred without notice, after which this version may no longer be accessible.}
\thanks{
 J. Hu and L. Song are with Department of Electronics, Peking University, China (e-mails: \{jingzhi.hu, lingyang.song\}@pku.edu.cn).
 }
\thanks{
 H. Zhang and H. Vincent Poor are with Department of Electrical Engineering, Princeton University, USA (e-mails: hongliang.zhang92@gmail.com, poor@princeton.edu).
 }
\thanks{
 K. Bian is with Department of Computer Science, Peking University, China (e-mail: bkg@pku.edu.cn).
 }
\thanks{Z. Han is with the Department of Electrical and Computer Engineering, University of Houston, USA, and also with the Department of Computer Science and Engineering, Kyung Hee University, South Korea (e-mail: zhan2@uh.edu).
}
}

\maketitle
\vspace{-3.5em}
\begin{abstract}
Semantic segmentation is a process of partitioning an image into multiple segments for recognizing humans and objects, which can be widely applied in scenarios such as healthcare and safety monitoring. To avoid privacy violation, using RF signals instead of an image for human and object recognition has gained increasing attention. However, human and object recognition by using RF signals is usually a passive signal collection and analysis process without changing the radio environment, and the recognition accuracy is restricted significantly by unwanted multi-path fading, and/or the limited number of independent channels between RF transceivers in uncontrollable radio environments. This paper introduces {\holosketch}, a novel RF-sensing system that performs semantic recognition and segmentation for humans and objects by making the radio environment reconfigurable. A metamaterial surface is incorporated into {\holosketch} and diversifies the information carried by RF signals. Using compressive sensing techniques, {\holosketch} reconstructs a point cloud consisting of the reflection coefficients of humans and objects at different spatial points, and recognizes the semantic meaning of the points by using symmetric multilayer perceptron groups. Our evaluation results show that {\holosketch} is capable of generating favorable radio environments and extracting exact point clouds, and labeling the semantic meaning of the points with an average error rate of less than $1\%$ in an indoor space.

\end{abstract}
\vspace{-1em}

\begin{IEEEkeywords}
RF sensing, reconfigurable intelligent surface, semantic segmentation, compressive sensing.
\end{IEEEkeywords}

\newpage

\section{Introduction}
%* 人体识别和语义分离是大家最近很感兴趣的领域。
%* 人体识别和语义分离大概说的是什么。
%* 传统的姿势识别和语义分离基于camera，但是camera涉及隐私，所以大家希望用RF信号进行语义分离和姿势成像，比如用WiFi进行成像。
%* 但是由于被动接受信道传播的问题，这样不好。
\IEEEPARstart{I}{n} computer vision, semantic segmentation seeks to partition the pixel set of an image into multiple subsets, with each subset having the same semantic meaning.
Owing to its wide applications in public safety and healthcare monitoring scenarios, %TODO 找到一两个文献，支持semantic在healthcare 和 monitoring上的应用。
semantic segmentation has garnered significant interest recently as a powerful tool for simultaneous recognition and localization of humans and objects.
Generally, semantic segmentation is conducted over images captured by video cameras and used to obtain meaningful representations for the images to simplify and facilitate further potential analyses~\cite{shapiro1992computer}.

% 已有 RF 系统
However, using video cameras to collect images for semantic segmentation inevitably introduces privacy concerns.
As a potential solution, recently using RF signals for profiling humans and objects has gained broad interest in research in this field.
Many RF-sensing systems based on WiFi signals or millimeter waves have been proposed for recognizing humans and objects~\cite{zhang2019Feasibility, jiang2018towards, Zhang2018CrossSense, Hsu2019Enabling, zhao2018through},
%zhang2019Feasibility mobicomm
%jiang2018towards mobicomm
%Zhang2018CrossSense mobicomm
%再找一篇mobicomm的
%已有3-5个工作的引用，mobicom，mobisys 为主，sigcomm, NSDI的工作为辅
or generating images that can be further used as materials for semantic segmentation~\cite{adib2015capturing, zhao2018rf, PedrossEngel2018Orthogonal}.
%zhao2018rf sigcomm
%PedrossEngel2018Orthogonal IEEE trans
%1-2个已有工作
However, human and object recognition by using RF signals usually employs a passive signal collection and analysis process without changing the radio environment.
Hence, due to the complicated and unpredictable radio environments, the accuracy and flexibility of recognition can be affected significantly by undesirable multi-path fading~\cite{honma2018Human, li2020programmable}, and/or the limited number of independent channels from the transmitters to the receivers in the conventional RF-sensing systems.

%* 通过主动改变信道传播环境，可以在接收信道处获得更多维度的信息。
%* 通过更多维度的信息，具有{\ris}的系统可以实现成像、语义识别等传统WiFi系统做不到的事情。
Recently, metamaterial surfaces, i.e., metasurfaces, have been developed as a promising solution to actively customize the undesirable propagation channels into favorable radio environments~\cite{Renzo2019Smart}.
% {\ris}的原理
A {\ris} is composed of a massive number of electrically controllable elements, which applies different phase-shifts to the signals reflected by it.
By optimally programming its elements, a {\ris} mounted in a radio environment can generate a massive number of independent propagation channels, which allows RF signals to carry diverse information about humans and objects.
Therefore, sensing based on the use of a {\ris} potentially lead to more accurate semantic recognition and segmentation results than those of traditional RF-sensing systems.
% 除去用更多、更复杂、更精密的射频收发装置，主动的改变信道环境可以有效的提升
This also introduces a new way of RF sensing that, instead of using more complicated and sophisticated RF transmitters and receivers, using a {\ris} with the capability of active channel customization to obtain high-fidelity results.

In this paper, we present {\holosketch}, a {\ris}-based RF-sensing system that can extract a point cloud from the RF signals reflected by humans and objects, which consists of the reflection coefficients in the spatial points, and perform semantic segmentation on the point cloud to recognize the humans and objects.
Specifically, via programming {\ris} configurations, {\holosketch} creates multiple independent propagation channels which facilitate the point cloud extraction.
Since {\ris}s rather than video cameras are employed for generating images for segmentation, {\holosketch} is designed to be privacy-protecting, and thus it has a wide variety of applications in healthcare and security monitoring scenarios.
% 由于文章中并没有展示加上纸板之后的情况，所以暂时不适用于加上下面这句话。
% Moreover, due to the penetrating capacity of the RF-signals, the developed {\holosketch} system can be used in logistics industry to check the object inside express boxes. 

The grand challenge of building a system that extracts point clouds without cameras is the absence of a method to directly capture an image of humans and objects by RF signals or to match an image to a certain set of received signals.
Therefore, existing systems cannot reconstruct images from RF signals directly.
%or an extracted point cloud (a set of data points in space) 
% 不可以先成像再获取点集，而是需要从提取的点集的基础上进行成像。
Instead, {\holosketch} seeks to extract the point clouds directly from processing the RF signals by compressive sensing, without relying on the reconstructed images.
% 这个东西从获取点集的角度讲特别像一个激光雷达， 虽然反而没有激光雷达的精确度。
The design of {\holosketch} is structured around three components that together provide an architecture for using compressive sensing and semantic segmentation for {\ris}-based RF-sensing systems:
%Each component serves a particular function as we describe below.
(1) A \emph{radio environment reconfiguration module} to create multiple independent propagation channels and facilitate compressive sensing based on {\ris},
(2) a \emph{point cloud extraction module} that extracts reflection characteristics of different positions in space, 
and (3) a \emph{semantic segmentation module} to recognize humans and objects and label the point clouds with their semantic meaning.

%{\holosketch} is implemented by using universal software radio peripheral~(USRP) devices and trained by data collected in indoor environment.
We evaluate the semantic segmentation capability of {\holosketch} over daily scenarios that involve a human and a set of practical objects.
Experiment results show that {\holosketch} can extract point clouds in space from RF signals and perform semantic segmentation accurately with an average error rate of less than $1\%$, given the setup of a human and three objects in a $1.6~m^3$ indoor space represented by $400$ evenly distributed points.
% With promising performance, {\holosketch} has proven the effectiveness of adopting {\ris} in RF-sensing system and would upend the way that RF sensing traditionally has been practiced.

The rest of the paper is organized as follows.
Section~\ref{sec: relate work} reviews related work on RF-sensing systems and video-image-based semantic segmentation.
Section~\ref{sec: preliminaries} provides preliminaries required to understand the design of {\holosketch}.
In Section~\ref{sec: overview}, we describe the system model {\holosketch}, including the system components and the coordination protocol among components.
Then, we describe the three component modules of {\holosketch}, i.e., the radio environment reconfiguration module in Section~\ref{RadioReconfiguration}, the point cloud extraction module in Section~\ref{Extraction}, and the semantic segmentation module in Section~\ref{Segmentation}.
Section~\ref{sec: system implementation} describes the implementation of {\holosketch}, and Section~\ref{sec: evaluation} shows the results of a performance evaluation of {\holosketch}.
Finally, we summarize the paper and discuss the technical issues for enhancing {\holosketch} in Section~\ref{sec: conclu}.

%%%%%%%%%%%%
\section{Related Work}
%%%%%%%%%%%%
\label{sec: relate work}

In this section, we first explain the advantages of the proposed {\ris}-based segmentation in this paper, then we summarize the related work for this paper, including the existing literature on RF-sensing systems and the image-based semantic segmentation technique.

One of the advantages of using the RF-sensing system for segmentation instead of image-based systems is for privacy protection.
Specifically, as RF-sensing can be independent of video cameras in data collection, which prevents any invasion of privacy.
% 这里要删掉啊。
%Some existing {\ris}-based RF-sensing systems collect training data through video images for supervised learning~\cite{Li2019Machine}, while {\holosketch} performs semantic segmentation on point clouds extracted by using compressive sensing without  raising privacy concerns due to video cameras.
%宋老师建议说这两个文章藏在后面。
%Some existing {\ris}-based RF-sensing systems collect training data through video images for supervised learning~\cite{Li2019Machine, Li2019Intelligent}, while {\holosketch} performs semantic segmentation on point clouds extracted by using compressive sensing without  raising privacy concerns due to video cameras.
In addition, regarding the customization of radio environments, the {\ris} is capable of reconfiguring the propagation channels between RF transceivers into various favorable shapes, and can be used to enhance the RF-sensing systems~\cite{li2020programmable}, e.g., the {\ris} can generate propagation channels which are mutually independent by programming its configurations.
Since RF signals traveling over independent channels generally carry more diverse information than those on coherent ones, adopting {\ris} in the RF-sensing system can potentially increase the accuracy of locating and recognizing humans and objects~\cite{Wang2016RF-Fall,Zhou2017Short}.
In other words, owing to the capability of reconfiguring radio environments, the {\ris}-based RF-sensing system can provide more reliable materials to perform semantic segmentation, compared with existing RF-sensing systems.

%======================
\subsection{RF-sensing Systems} 
%======================
Recent years have witnessed much interest in RF-sensing systems for human and object recognition.
Existing systems work by analyzing the influence of human body and objects on the RF signals.
Different systems are designed for people localization~\cite{adib2015multi,patwari2010rf} and particular posture- and gesture-identification~\cite{Kellogg2014Bringing,sigg2013rf,lien2016soli,Wang2016RF-Fall, tian2018rf}.
Besides, RF-sensing also proves to be feasible for imaging humans and objects with the help of MIMO techniques~\cite{adib2015capturing, Gollub2017Large}.
%TODO forV6 总感觉这个related work部分并没有写完，需要看一下别人的related work 需要写到什么程度。

%==================================
\subsection{Image-based Semantic Segmentation} 
%==================================
In semantic segmentation, 
%is to partition a set of pixels into multiple semantic segments.
usually, each segment is a set of pixels of the image which collectively represent one semantically meaningful object, e.g., a human or a suitcase in the image.
Most semantic segmentation approaches are based on processing images captured from $2$D video cameras.
The $2$D semantic segmentation techniques are becoming mature due to the increasing availability of the deep convolutional network model, e.g., fully convolutional networks~\cite{Long_2015_CVPR}, R-CNNs~\cite{ren2015faster}, and deep convolutional encoder-decoders~\cite{Badrinarayanan2017SegNet}.

Different from the increasingly matured segmentation in $2$D, the semantic segmentation in $3$D has gained more attention.
Authors in~\cite{mehta2017vnect} proposed to construct human skeletons in $3$D from $2$D images by using convolutional neural networks combined with kinematic skeleton fitting.
Besides, authors in~\cite{qi2017pointnet} explored deep learning architectures capable of reasoning about geometric data other than $2$D images such as $3$D point clouds and meshes.
This lays the foundation for semantic recognition and segmentation in {\ris}-based RF-sensing systems where the reflection coefficients in space points are captured and form a point cloud.

%%%%%%%%%%%%
\section{Preliminaries}
%%%%%%%%%%%%
\label{sec: preliminaries}
In this section, we show preliminary results that lead to successfully building the three components of {\holosketch}. 
%first explain show the {\ris} customizes radio environments through a pilot experiment. Then, we demonstrate the feasibility of extracting point cloud via compressive sensing methods. Finally, we introduce the multilayer perceptron~(MLP) for semantic recognition and segmentation of point clouds.

%==================================
\subsection{Pilot Experiment for Reconfiguring Radio Environment by {\Ris}}
%==================================
\label{ssec: ris model}

\begin{figure}[!t] 
	\center{\includegraphics[width=0.55\linewidth]{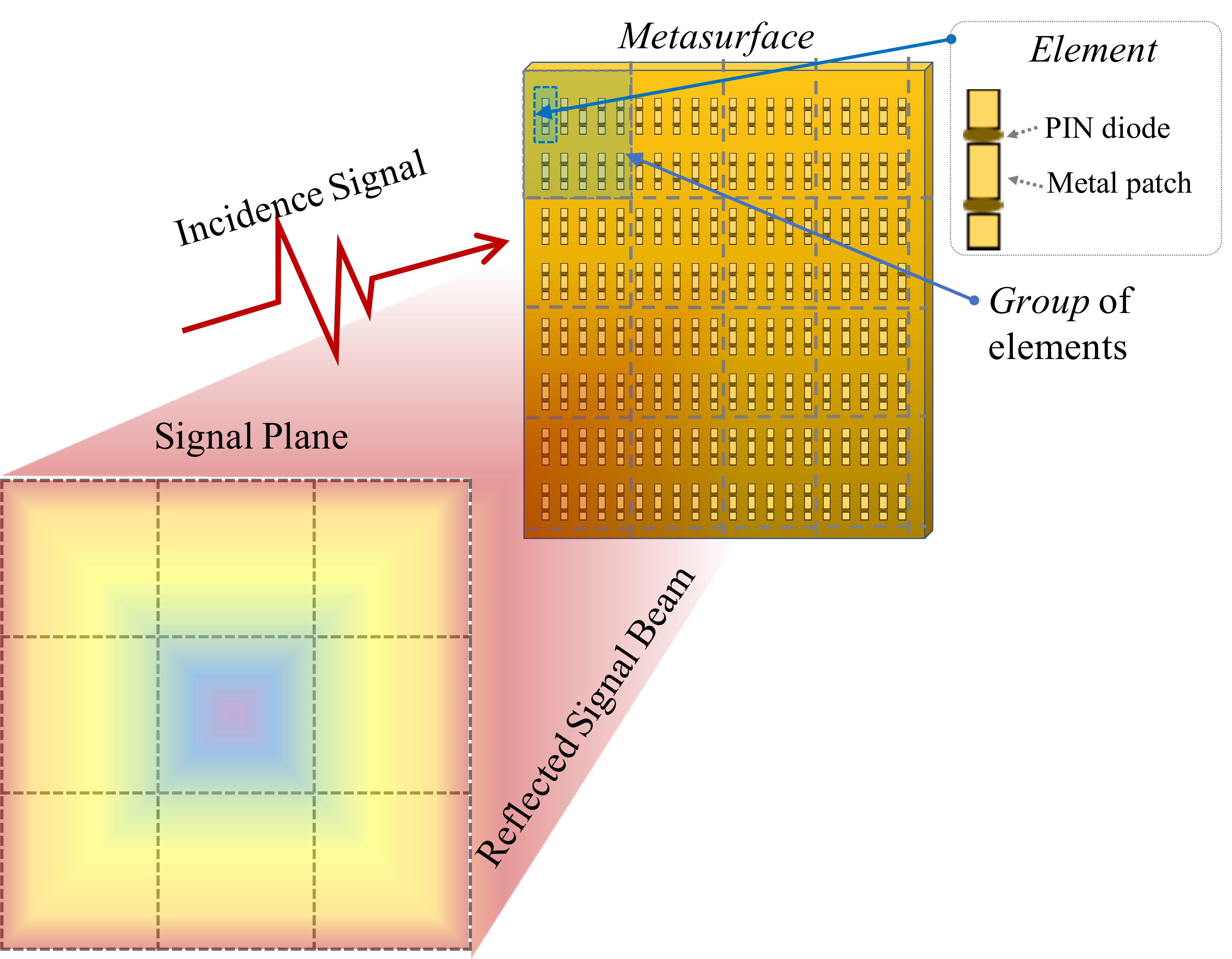}}
	\vspace{-1em}
	\caption{{\Ris} elements and signal reflection on {\ris}.}
	\label{fig: RISExample}
\end{figure}

\begin{figure}[!t]
	\center{\includegraphics[width=0.8\linewidth]{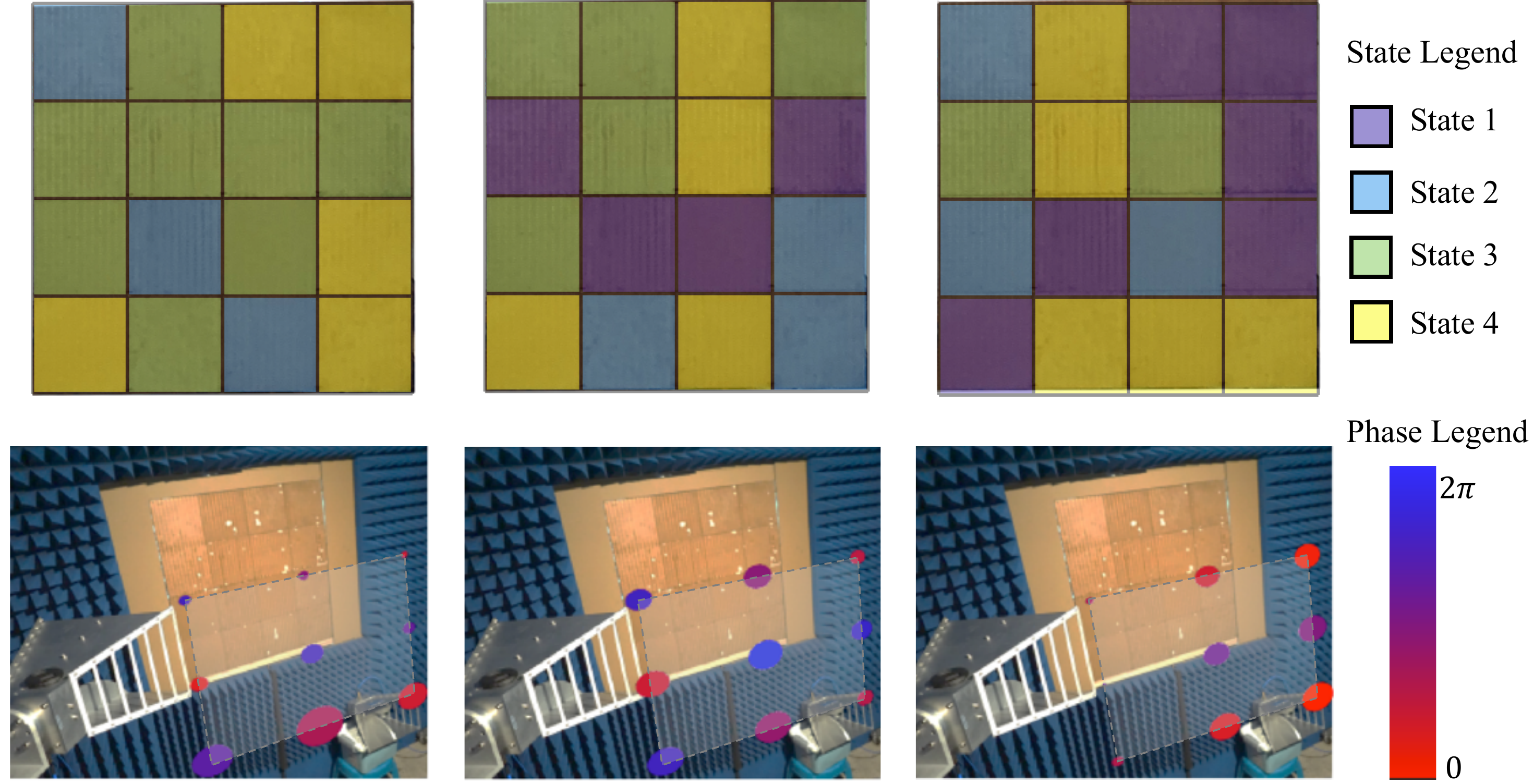}}
	\vspace{-1em}
	\caption{Configurations of the {\ris} and the corresponding reflected signals at different positions.}
	\label{fig: radio environment reconfiguration}
\end{figure}

We first explain the {\ris} and its elements, and then elaborate on a pilot experiment which illustrate their capability of reconfiguring radio environments.

%----------------------------------------------
\subsubsection{{\Ris} elements and states}
%----------------------------------------------
{\Ris} is an artificial thin film of reconfigurable electromagnetic materials, which is composed of a massive number of uniformly distributed \emph{{\ris} elements}.
As shown in Fig.~\ref{fig: RISExample}, {\ris} elements are arranged in a two-dimensional array.
For each {\ris} element, it can adjust its response to the incident RF signals by leveraging positive-intrinsic-negative~(PIN) diodes~\cite{cui2010metamaterials}.
We refer to the different responses of a {\ris} as the \emph{states} of the {\ris} element.
The {\ris} elements can be set to different states, and each state of the {\ris} element shows its own electrical property, leading to a unique reflection coefficient for the incident RF signals.
To be specific, the reflection coefficients of a {\ris} element at different states can be represented as a complex number.
The amplitude and phase of the reflection coefficient denote the amplitude ratio and the phase shift between the reflected and incident signals, respectively.

Due to the number of {\ris} elements within a {\ris} is usually large, it is costly and inefficient to control each {\ris} element independently.
To reduce the controlling complexity, we divide the {\ris} elements into multiple \emph{groups}, as shown in Fig.~\ref{fig: RISExample}.
Therefore, the states of all {\ris} elements can be represented by the states of the groups, which we referred to as the \emph{configuration} of the {\ris}.
% 可以在{\ris}的介绍中都不用具体的数学，到compressive sensing中再用。

%TODO 要在这其中体现出{\ris}的数学原理。
%----------------------------------------------------------
\subsubsection{Visualization of reconfiguring radio environment}
%----------------------------------------------------------
Through changing its configuration, the {\ris} is able to modify the waveforms of the reflected signals and form directional beams~\cite{di2019hybrid}.
Based on~\cite{tang2019wireless}, the beamforming capability of a {\ris} can be characterized as follows.
Given a spatial position in front of the {\ris}, the radio signal at that position is
\begin{equation}
\label{equ: math of RIS}
y = \sum_{n=1}^N\frac{\lambda \exp(-j2\pi d_n/\lambda)}{4\pi d_n}\cdot r_n(s_n)\cdot x_n
\end{equation}
where 
$N$ denotes the number of {\ris} elements,
$\lambda$ is the wavelength of the transmitted sine wave, 
$n$ denotes the index of a {\ris} element,
$d_n$ denotes the distance from the it to the position,
$s_n$ denotes the state of it
and $x_n$ is the incident signal of it.
Besides, $r_n(s_n)$ is the reflection coefficient of the {\ris} element $n$, which is dependent on state $s_n$ and the physical structure of the {\ris} element.

With the beamforming capability, the {\ris} can reconfigure the radio environment for the incident signals and reflect diverse beam patterns in front of it.
% 是怎么做实验的
To show the {\ris}'s capability of reconfiguring the radio environment we conduct a pilot environment by changing the configuration of the {\ris} randomly, and measuring the reflected signals at $9$ different positions on a plane around $1.2$~m in front of the {\ris}.
% 是怎么展示结果的。
In Fig.~\ref{fig: radio environment reconfiguration}, the configurations of the {\ris} are depicted in the upper part, where the four colors indicate four different states of groups.
Besides, the corresponding reflected signals are visualized as colored solid circles~(red, blue, or purple) in the lower part.
In the lower part, the size of each circle is proportional to the signal amplitude, and the color represents the signal phase.
It can be observed that by changing the configuration of {\ris}, the strength and phase of the reflected signals at different positions can be modified.

%========================
\subsection{Feasibility of Extracting Point Cloud via Compressive Sensing}
%========================
\label{ssec: pce from rf signals}

\begin{figure}[!t]
	\center{\includegraphics[width=0.6\linewidth]{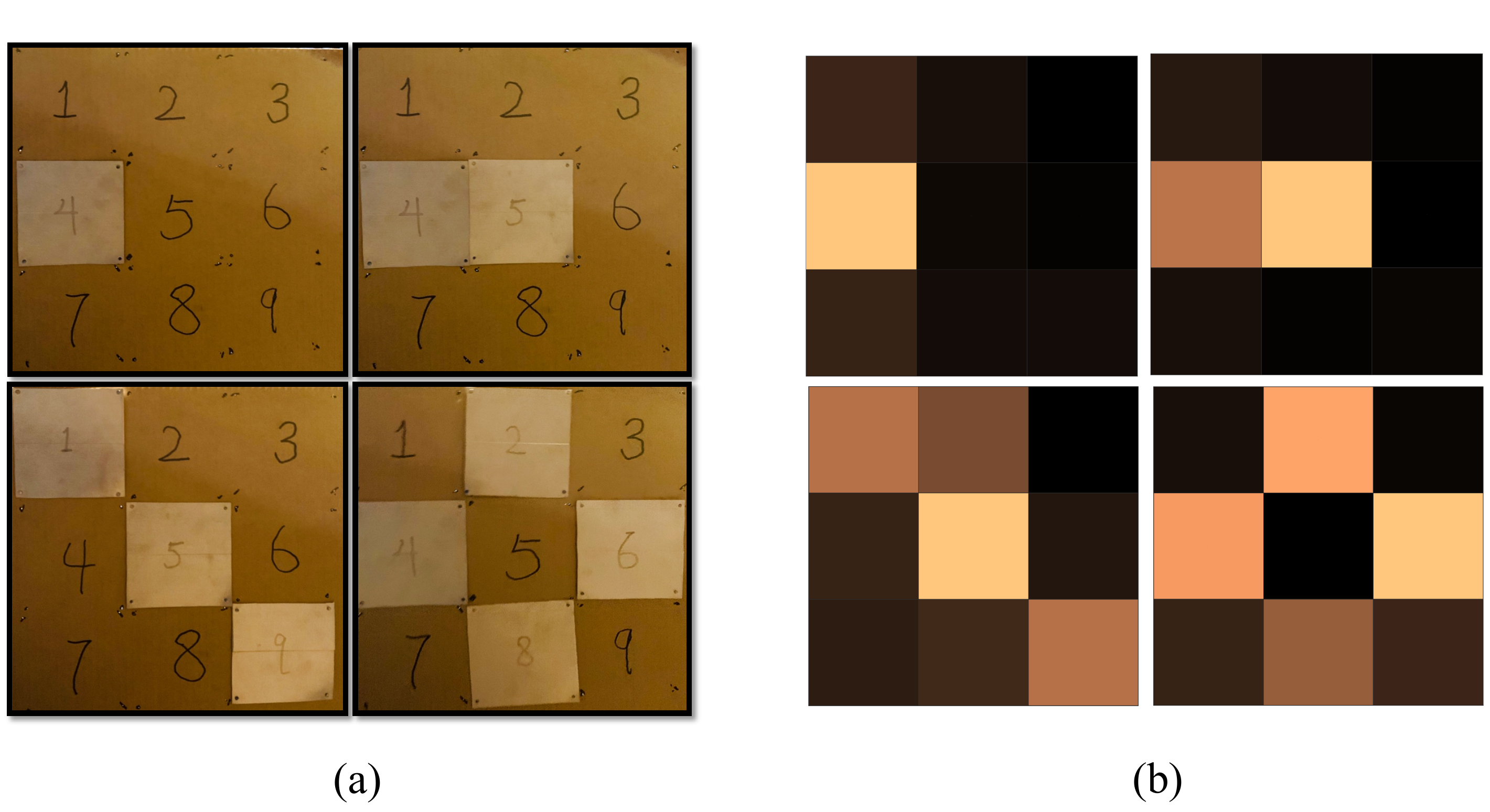}}
	\vspace{-1em}
	\caption{(a) The signal plane with $10$~cm $\times$ $10$~cm metal patches at part of the $9$ positions; and (b) illustrations of the reconstructed absolute values of the reflection coefficients at different positions by solving~(\ref{prob: l1 norm minimization}).}
	\label{fig: reconstructed figure}
\end{figure}

In the following, we demonstrate the method of extracting point cloud based on the received RF signals in a {\ris}-assisted system.
Specifically, we first introduce the compressive sensing technique, and then illustrate the feasibility of recovering the reflection coefficients using the compressive sensing technique.
%----------------------------------------------------------
\subsubsection{Compressive sensing technique}
%----------------------------------------------------------
The compressive sensing technique is developed to recover signals from highly incomplete information~\cite{han_COMPRESSIVE}.
% 由于我们要从有限的测量信号中还原出众多空间点处的反射系数，所以需要利用compressive sensing。
We need to restore the reflection coefficients at many positions in space from a limited amount of measurement signals and extract the point clouds.
%, and thus the compressive sensing technique is used.
Hence, when we use the compressive sensing technique, the main focus is to solve some target variable $\bm x$ in underdetermined equation under some noise $\bm e$ given measurement $\bm y$ and measurement matrix $\bm H$, i.e.,
\beq
\label{equ: target func in compressive sensing}
\bm y = \bm H \bm x + \bm e,
\eeq
where the dimension of $\bm y$, i.e., $\dim(\bm y)$, is much less than $\dim(\bm x)$, and the number of non-zero elements of $\bm x$ is much less than its dimension, indicating that $\bm x$ is \emph{sparse}.
Mathematically, the sparsity can be evaluated by $l_0$-norm defined by $\|x\|_0 = \left|\{ i\in[1,\mathrm{dim}(\bm x)] | x_i \neq 0  \}\right|$, where $|\cdot|$ denotes the number of elements in the contained set.

Given $\bm x$ to be sparse, solving $\bm x$ in~(\ref{equ: target func in compressive sensing}) can be done by minimizing $\|\bm x\|_0$ subject to $\|\bm y - \bm H \bm x\|_2\leq \epsilon$, where $\epsilon $ denotes the variance of noise $\bm e$.
However, due to the non-convexity of $l_0$ norm minimization, it is infeasible for most practical applications.
An alternative approach suggested in~\cite{candes2006stable} is the $l_1$-norm minimization approach, which solve sparse vector $\bm x$ in~(\ref{equ: target func in compressive sensing}) by solving
\beq
\label{prob: l1 norm minimization}
\hat{\bm x} = \arg\min_{\bm x} \|\bm x\|_1 \quad \text{s.t.}~\| \bm H \bm x - \bm y\|_2\leq \epsilon.
\eeq
As shown in~\cite{candes2006stable}, if a sparse variable $\bm x$ exists such that $\bm y = \bm H \bm x+ \bm e$, for some small error term $\|\bm e\|_2\leq \epsilon$, then the solution for (\ref{prob: l1 norm minimization}) $\hat{\bm x}$ will be close to the real $\bm x$.

%----------------------------------------------------------------------
\subsubsection{Feasibility of recovering the reflection coefficients}
%----------------------------------------------------------------------
As an example to verify the feasibility of extracting point cloud via compressive sensing, we try using {\ris} {\holosketch} to reconstruct the average reflection coefficients of a $9$-rectangle grids in space.
% 实验部分说明
We place one $10$~cm $\times$ $10$~cm metal patch at the $9$ positions on the signal plane shown in Fig.~\ref{fig: radio environment reconfiguration} in turn and measure the corresponding received signals when the {\ris} takes $5$ random configurations.
Based on the received signal sequences at the $9$ points, the measurement matrix $\bm H$ is obtained, i.e.,
\beq
\bm H = \begin{bmatrix}
\hat{\bm y}_1 & \dots & \hat{\bm y}_9
\end{bmatrix}.
\eeq
Here, $\hat{\bm y}_i = \bm y_i^{M}\!-\!\bm y^{B}$~($i\!\in\![1,9]$) is the $i$-th column of $\bm H$ and indicates the influence of a metal patch at $i$-th grid on the received signals, where $\bm y_i^M$ denotes the received signal vector given the metal patch located at the $i$-th grid and $\bm y^B$ indicates the received signal vector given no metal patch presented.

Then, we place multiple metal patches on the $9$ grids, obtain received signal $\bm y$, and solve (\ref{prob: l1 norm minimization}) to obtain $\bm x$, which is a $9$-dim complex vector indicating the average reflection coefficients of the $9$ grids with respect to the metal patch.
Fig.~\ref{fig: reconstructed figure}~(a) shows the photo of the pattern that the metal patches form, where the light~(yellow) regions are the metal patches, and the dark~(brown) regions are the cardboards, which have a negligible impact on wireless signals.
Fig.~\ref{fig: reconstructed figure}~(b) illustrates the reconstructed amplitude of the average reflection coefficients at the $9$ grids, where the lighter color indicates a higher reflection coefficient value indicating the position of metal patches.
Comparing Figs.~\ref{fig: reconstructed figure}~(a) and~(b), we can observe that solving (\ref{prob: l1 norm minimization}) successfully reconstructs the reflection characteristics of the $9$ grids, which shows the feasibility of extracting point cloud via compressive sensing based on {\ris}.

% HJZ 将related work那个文献藏在了这里。
Based on the extracted point clouds, {\holosketch} can then perform semantic recognition and segmentation on the points.
Compared with the existing RF-sensing system with a {\ris}~\cite{Li2019Machine} which relies on cameras to capture profiles of human and links them with RF signals, {\holosketch} uses point clouds extracted by using compressive sensing as input data, which prevents raising privacy concerns.

%=========================
\subsection{Point Clouds Recognition using Symmetric Multilayer Perceptron Groups}
%=========================
\label{ssec: mlp}
In order to recognize humans and objects in an extracted point cloud, semantic segmentation is necessary for feature learning~\cite{Long_2015_CVPR,dai2015boxsup}.
% (Given a set of 3D points $\{p_i\}$, the aim in our work is to generate some part regions corresponding to our semantic subcategories from these 3D points, which can be achieved by neural network.)
To be specific, given a set of points, we aim to label each point with its semantic meaning.
% 点集合数据的特点
However, as the point clouds are essentially point sets, they are invariant to changing order, which make them different from conventional data structures for semantic segmentation such as pixel images~\cite{qi2017pointnet}.
Therefore, traditional semantic segmentation methods based on convolutional neural networks~\cite{Long_2015_CVPR, dai2015boxsup} %这里举一两个CNN based semantic segmentation的文献
cannot be used here, as they rely on the convolution operation on input data arranged in spatial order for regional feature extraction.
In order to handle the unordered properties of point cloud data structure, the input data need to be treated by symmetric functions.
As proven in~\cite{qi2017pointnet}, this can be achieved effectively by using multi-layer perceptons~(MLPs) with shared parameters to treat the feature vectors of points in the point cloud.

\begin{figure}[!t] 
    \center{\includegraphics[width=0.5\linewidth]{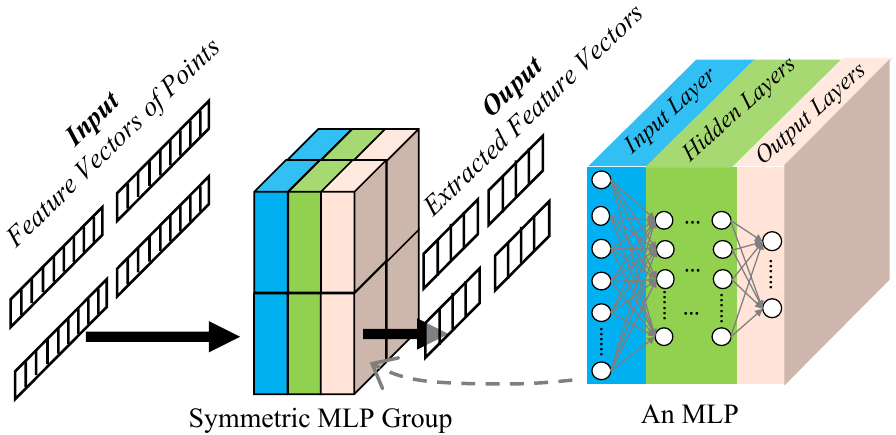}}
    \vspace{-1em}
    \caption{Illustration on a symmetric MLP group to process the feature vectors in a point cloud.}
    \label{fig: mlp}
\end{figure}

An MLP can be considered as a mathematical function mapping that is constituted by many tunable functions in simpler forms~\cite{goodfellow2016deep}. 
In {\holosketch}, the MLP is used to extract a more representative feature vector of each point from its ordinary one, in order to facilitate point recognition.

More specifically, as shown in Fig.~\ref{fig: mlp}, an MLP contains multiple layers of \emph{neurons} that extract information from the input feature vector.
Each neuron takes input from the connected neurons in the lower layer, handles them by weighted summation with bias and an activation function~(e.g., sigmoid), and outputs the result value to the upper layer.
% 为了节省地方删去了如下内容
%Denote the $i$-th neuron at the $n$-th layer by Neuron $n-i$, and the output value of it can be expressed as 
%\beq
%a_i^{n} = \sigma(\sum_j w^{n}_{ij}a_j^{n-1} - b^{n}_i),
%\eeq
%where $a_j^{n-1}$ indicates the output value of the Neuron $(n-1)-j$, $w^n_{ij}$ denotes the connection weight between Neuron $(n-1)-j$ and Neuron $n-i$, $b^n_i$ denotes the bias of Neuron $n-i$.
%Parameters $b_i^n$ and $w^n_{ij}$ for all the neurons and connections are referred to the parameter of MLP.

In {\holosketch}, since the point cloud is an unordered set, the MLP to treat each point needs to be symmetric.
Therefore, we ensure that the MLPs to have the same parameters by parameter sharing, and refer to them as a symmetric MLP group.
In a symmetric MLP group in {\holosketch}, the input to each MLP can be the position and reflection coefficient of its point as well as an extracted feature vector from a previous MLP, and the output can be a feature vector with higher/lower dimensions or the semantic recognition result for that point.

%%%%%%%%%%%%%%
\section{System Model}
%%%%%%%%%%%%%%
\label{sec: overview}

In this section, we describe the system model of {\holosketch} by describing three system components and a protocol that coordinates the messages transmission among the components. 

%========================
\subsection{System Components}
%========================
\begin{figure}[!t] 
	\center{\includegraphics[width=0.5\linewidth]{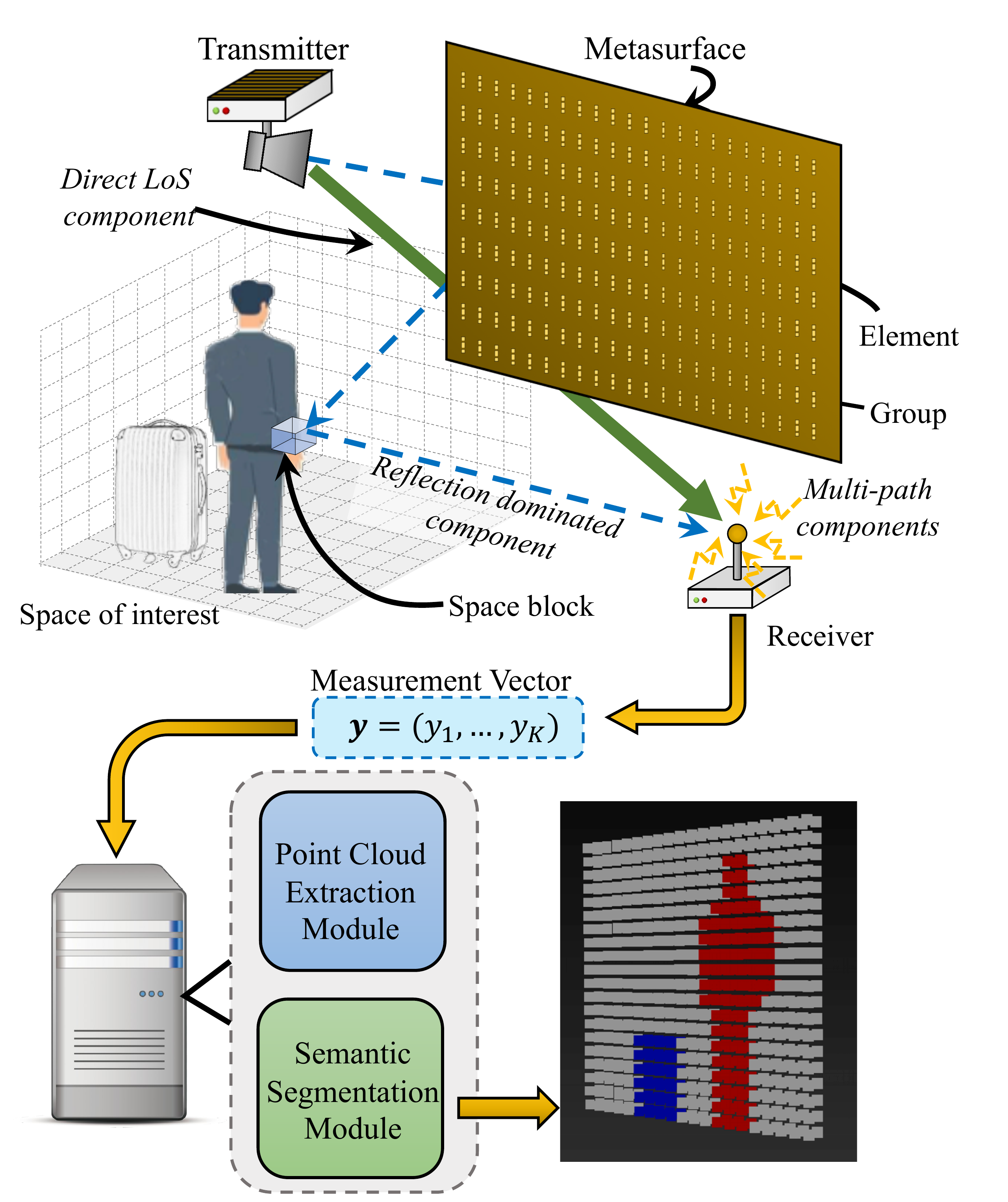}}
	\vspace{-1em}
	\caption{Components of {\holosketch}: {\ris}-based radio environment reconfiguration~(as shown in the top half of the figure), point cloud extraction, and semantic segmentation modules~(as shown in the bottom half of the figure).}
	\label{fig: pce and ss}
	%TODO (Maybe) 要把上面的部分用一个灰色底色框包成为Radio Environment Reconfiguration Module.
\end{figure}

{\holosketch} is an RF-sensing system that can extract point clouds of humans and objects in space and perform semantic segmentation on the point clouds.
As illustrated in Fig.~\ref{fig: pce and ss}, the system contains the following three component modules:

\begin{itemize}[leftmargin=*]
\item \textbf{Radio environment reconfiguration module}: 
    This module contains a pair of RF transceivers and a {\ris}.
    The pair of transceiver consists of a transmitter~(Tx) and a receiver~(Rx), which are equipped with single antennas to transmit and receive RF signals, respectively. 
    The {\ris} reflects and reshapes incident signals according to its configuration.
    The transmitted signals from the Tx are modified by the {\ris} and then reach the humans and objects, carrying out the information of them to the Rx.
\item \textbf{Point cloud extraction module}: 
    This module is implemented in the sever connected to the Rx and adopts a compressive sensing technique to extract the point cloud from the baseband RF signals from the Rx, which is the reflection coefficients at different space points.
\item \textbf{Semantic segmentation module}: 
    The obtained point cloud in the previous module is then inputted into the semantic segmentation module implemented in the server.
    The semantic segmentation module recognizes humans and objects in the point cloud and labels each point with its semantic meaning.
    In this module, the MLPs are adopted and trained by supervised learning method.
\end{itemize}

%======================
\subsection{Coordination Protocol}
%======================
\label{ssec: operating protocol}
\begin{figure}[!t] 
\center{\includegraphics[width=0.5\linewidth]{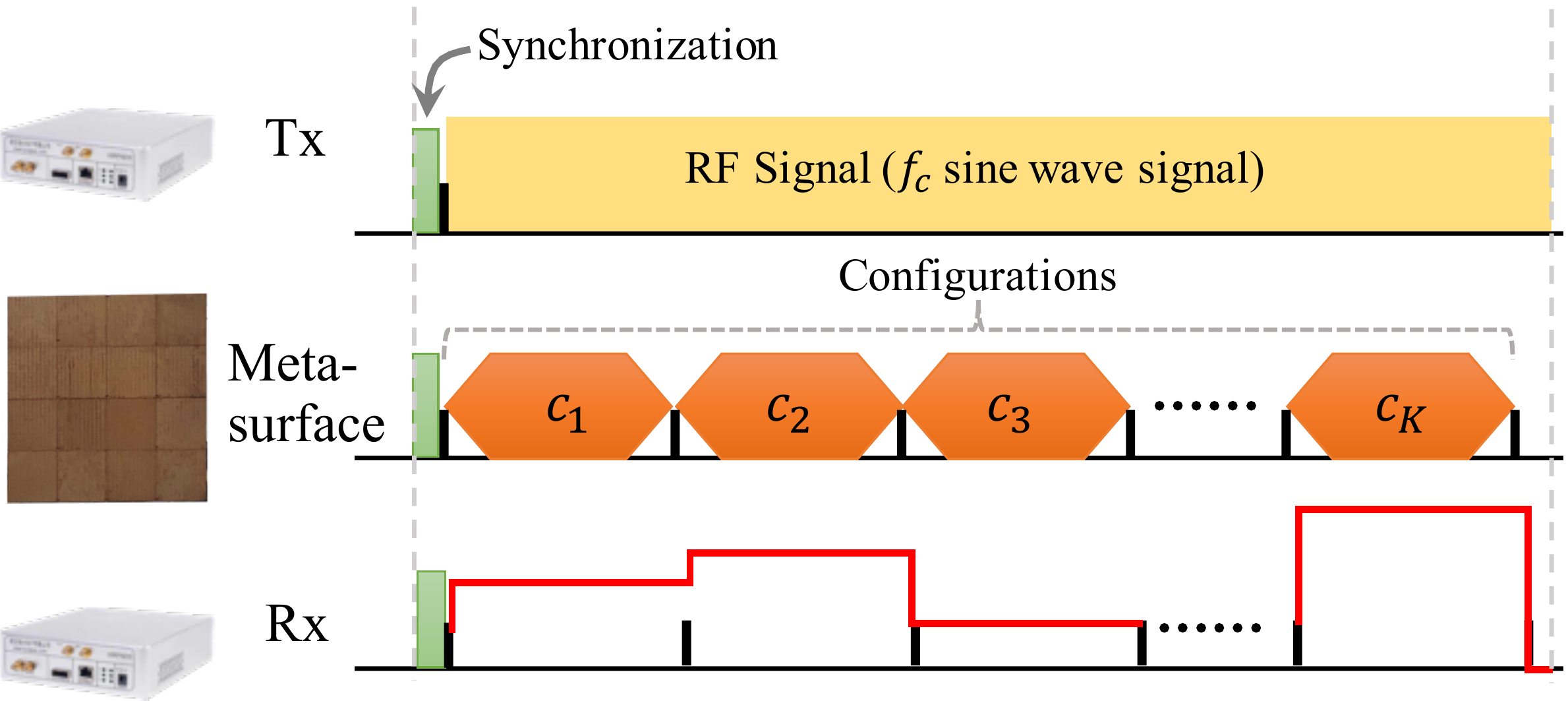}}
	\vspace{-0.5em}
	\caption{Illustration on the data collection phase.}
	\label{fig: protocol}
\end{figure}

In the following, we propose a protocol to coordinate the component modules of {\holosketch} to perform RF-sensing, point cloud extraction, and semantic segmentation.
In the protocol, the timeline is slotted and divided into \emph{cycles}, and {\holosketch} operates in a synchronized and periodic manner.
Each cycle is constituted of two phases: \emph{data collection} and \emph{signal processing} phases. 
As illustrated in Fig.~\ref{fig: protocol}, in the data collection phase, the {\ris} changes configuration sequentially.
The receiver measures the received signals during every configuration and stores them as a vector, referred to as the \emph{measurement vector}.

The signal processing phase follows the data collection phase, where the point cloud extraction and semantic segmentation modules are invoked to process the obtained data sequentially.
In the following part, we describe the data collection and signal processing phases in detail.

%-------------------------------------------
\subsubsection{Data collection phase}
%-------------------------------------------
%In the data collection phase, the radio environment reconfiguration module uses RF signals to sense the human and object in the space and generate measurement vectors.
At the beginning of the data collection phase, the Tx first transmits a starting signal to the {\ris} and the receiver for synchronization.
Then, the Tx starts to transmit sine wave signal with frequency $f_c$, and the {\ris} changes sequentially from the first to the $K$-th configuration, which are denoted by $\bm c_1$ to $\bm c_K$ with $K$ being the total number of configurations, as shown in Fig.~\ref{fig: protocol}.
Specifically, $\bm c_k$~($k\in[1,K]$) is a $L$-dim vector where $L$ is the number of groups of the {\ris}.
Moreover, the $K$ configurations of the {\ris} constitutes \emph{measurement matrix} $\bm C$, i.e., $\bm C = (c_1,...,c_K)$.
While random $\bm C$ can be adopted, we propose the method to obtain an optimized configuration matrix in Section~\ref{RadioReconfiguration}.

At the end of the data collection phase, the Rx generates the $K$-dim measurement vector $\bm y$ by taking the averages of the received signals within $K$ configuration duration.
%For example, $y_k$, the $k$-th element of measurement vector $\bm y$, is the average of the received signals when {\ris} is at the $k$-th configuration.
Then, the Rx sends $\bm y$ to the sever for point cloud extraction and semantic segmentation.

%-------------------------------------------
\subsubsection{Signal processing phase}
%-------------------------------------------
After receiving the measurement vector generated by the Rx, the sever first invokes the point cloud extraction module to extract the point cloud of the humans and objects from the measurement vector.
Then, the generated point cloud image is processed by the semantic segmentation module, which provides each point with the label representing its semantic meaning.
The point cloud extraction module will be introduced in Section~\ref{Extraction}, and the semantic segmentation module will be elaborated in Section~\ref{Segmentation}.

%%%%%%%%%%%%%%%%%%%%
\section{Problem Formulation and Algorithm Design of {\holosketch}}
%%%%%%%%%%%%%%%%%%%%

In this section, we describe the problem formulation and algorithm design of {\holosketch}'s three core components,, which enables the {\holosketch} to perform RF segmentation accurately.

%=================================
\subsection{Radio Environment Reconfiguration}
%=================================
\label{RadioReconfiguration}
In this section, we describe how the radio environment reconfiguration module of {\holosketch} derives its configuration matrix.
While random configuration matrices are available, to facilitate the point cloud extraction and semantic segmentation requires the configuration matrix to be optimized.
As the point clouds are supposed to be extracted from measurement vector $\bm y$ in each cycle, we consider selecting the configuration matrix with which $\bm y$ can carry the largest amount of information about the humans and objects.

The information about humans and objects is contained in the reflection coefficients at different positions in space.
Specifically, the reflection coefficients can be expressed as a $M$-dim vector $\bm \eta$, where $M$ is the cardinality of a set of pre-assigned spatial positions whose reflection coefficients we want to restore.
The $j$-th element of $\bm \eta$ indicates the average reflection coefficient around the $j$-th pre-assigned spatial position.
Consequently, we need to optimize $\bm C$ so that $\bm \eta$ can be restored from $\bm y$ with the highest accuracy.

In {\holosketch}, though the number of spatial positions, $M$, can be large, most spatial positions are empty, and thus tend to have zero reflection coefficients.
Besides, for the spatial positions where the humans and objects reside, only those which are around the object surfaces with specific angles can reflect the incident signals towards the receiver and have non-zero reflection coefficients.
Therefore, $\bm \eta$ is a sparse vector and can be solved by using the compressive sensing technique, which we introduced in Section~\ref{ssec: pce from rf signals}.

Based on~\cite{elad2007optimized}, to minimize the loss between the reconstructed $\bm \eta$ and the actual one, we can minimize the average mutual coherence~(AMC) of $\bm H$, which is defined as
\beq
\label{equ: mutual coherence}
\mu(\bm H)= \frac{1}{M(M-1)}\sum_{m,m'\in[1,M],m\neq m'}\frac{|\bm h_m^T \bm h_{m'}|}{\|\bm h_m\|_2\cdot \|\bm h_{m'}\|_2}.
\eeq
Here, $\bm h_m\in \mathbb C^K$ and $\bm h_m$ is the $m$-th column of $\bm H$.
The $i$-th element of $\bm h_m$~($i\in[1,K]$) indicates the influence of a surface with reflection coefficient $1$ at the $m$-th position on the received signals, given the {\ris} is at the $k$-th configuration.
Measurement matrix $\bm H$ is determined by $\bm C$, and we can obtain the value of $\bm H = \bm g(\bm C)$ according to Appendix~A.

Based on~(\ref{equ: mutual coherence}), we can formulate the optimization for radio environment reconfiguration the following mutual coherence minimization problem:
\begin{align}
(\text{P1})~
\min_{\bm C}~
% min
& \mu(\bm H),  \\
% st
s.t. \quad 
%const 1
&\bm H = \bm g(\bm C), \nonumber \\
%const 2
\label{opt: reform 6 st2}
&c_{k,l}\in[1,N_s]~\forall k\in[1,K], l\in[1,L], \nonumber
 \end{align}
where $N_s$ denotes the number of states of a {\ris} element and $c_{k,l}$ denotes the state of the $l$-th group in the $k$-th configuration.
To solve (P1), we propose the configuration optimization algorithm, which is summarized in Algorithm~1.

\begin{algorithm}[!t]  
\label{alg: config design alg}
\small
\caption{Configuration Optimization Algorithm}
	\SetKwInOut{Input}{Input}
	\SetKwInOut{Output}{Output}
\Input{Initial random configuration matrix $\bm C^{(0)}$; Initial population size for genetic algorithm~(GA) $N_P$.}
\Output{Optimal AMC $\mu^*$ and configuration matrix $\bm C^*$.}

Set $\bm C^* = \bm C^{(0)}$, and compute initial $ \mu^*$ based on~(\ref{equ: mutual coherence}) and Appendix~A given $\bm C$\;

Set the number of consecutive iterations with no improvements as $N_{\mathrm{non}} = 0$ and current frame index $k=1$\;

\While{True}
{
    Based on Appendix~A, generate continuous configuration matrix $\tilde{\bm D}$ based on $\bm C^*$, and denote the $k$-th row of $\tilde{\bm D}$ as $\tilde{\bm d}_k$ and the other rows as $\bm \tilde{\bm D}_{-k}$\;

    Invoke pattern search algorithm~\cite{lewis2007implementing} to solve $\tilde{\bm d}_k^* = \arg\min \mu([\tilde{\bm d}_k, \bm \tilde{\bm D}_{-k}]\bm A )) $, where $\bm A$ is defined in Appendix~A\;

    Round up $\tilde{\bm d}_k^*$ to discrete configuration vector $\bm c_k'$ by 
    $(\bm c_{k}')_l = \arg\max_{j\in[1,N_s]}(( \tilde{\bm d}_k )_{(L-1)N_s+j})$\;

    % 我觉得可以把P1换一种表示形式，自变量表示为c_1...c_K
    Invoke genetic algorithm~\cite{goldberg2006genetic} to solve $\bm c_{k}^* = \arg\max_{c_{k,l}\in[1,N_s]} \mu(\bm g(\bm c_k, \bm C^*_{-k}))$, with the initial population consisting of $\bm c_{k}'$ and $(N_P-1)$ random configurations, and denote the result AMC as $\mu^{*'}$\;
    
    If $\mu^{*'}<\mu^*$, update $\mu^{*} =\mu^{*'}$, the $k$-th row of $\bm C^*$ to be $\bm c_{k}^*$; otherwise, set $N_{\mathrm{non}} = N_{\mathrm{non}} + 1$\;

    If $N_{\mathrm{non}} < K$, set $k = \mathrm{mod}(k+1, K) + 1$; otherwise, return $\mu^*$ and $\bm C^*$\;
}
\end{algorithm}

%%%%%%%%%%%%%%%%%%%%%
\subsection{Point Clouds Extraction}
%%%%%%%%%%%%%%%%%%%%%
\label{Extraction}

%In this section, we describe the point clouds extraction module and the compressive sensing technique in {\holosketch}.

%-----------------------------------------------------------
\subsubsection{Measurement Matrix Construction}
%-----------------------------------------------------------
\label{sec: obtain H}

To perform point cloud extraction, we need first to construct the accurate values of the measurement matrix given the optimized configuration matrix $\bm C^*$ obtained by solving~(P1), as mapping $\bm g$ in Appendix~A does not take the influence of the environment into account.
% 得解释清楚我们为什么要这么做
%The process to construct $\bm H$ is illustrated in Fig.~\ref{}.
To be specific, we first assign a set of $M$ spatial positions, which is denoted by $\mathcal M = \{(x_m, y_m, z_m)|m\in[1,M]\}$.
Then, we position a metal patch at the center of each spatial position, and collect the measurement vectors using the protocol described in Section~\ref{ssec: operating protocol} with the {\ris} using optimized configuration matrix $\bm C^*$.
When the metal patch is at the $m$-th~($m\in[1,M]$) spatial position, the collected measurement vector is denoted as $\hat{\bm y}_{\bm C^*, m}$.
%Following that, we collect the measurement vector when no metal patch is placed, i.e., the measurement vector for the background reflection, which is denoted as $\bm y^B$. 这种方法下面好像不需要收集background信号，两个一减就可以了。
Then, we remove the metal patch and obtain the measurement vector accounting for the background scattering, which is denoted as $\bm y^B_{C^*}$.
Based on the superposition property of the wireless signals, the channel gain for the propagation channel reflected at the $m$-th space block can be obtained by 
\beq
\bm h_{m}^* = ({\hat{\bm y}_{\bm C^*, m} - \bm y^B_{C^*}})/{\eta^{M}_{m}},
\eeq
where $\eta^{M}_{m}$ is the reflection coefficient of the metal patch at the $m$-th space block towards the Rx antenna. 
Moreover, for normalization, we assume $\eta^{M}_{m} = 1$, and thus each element in obtained $\bm\eta$ indicates the reflection coefficient with respect to the metal patch at that spatial position. 

Then, the measurement matrix $\bm H^* \in \mathbb C^{K\times M}$ corresponding to $\bm C^*$ can be formed by $\bm H ^*= [\bm h_1^*,...\bm h_M^*]$.
We use $\bm H$ to extract point cloud of target objects.
Denote the collected measurement vector as $\bm y$, and then the following equation holds
\beq
\label{equ: practical compressive sensing}
\bm y - \bm y^B  = \bm H \tilde{\bm \eta} + \bm e,
\eeq
where $\bm \eta\in \mathbb C^M$ is the reflection coefficient vector of the pre-assigned spatial positions with respect to metal patches.
Then, with known $\bm y$, $\bm y^B$, and $\bm H$, we can extract the point cloud by solving~(\ref{equ: practical compressive sensing}) to obtain $\bm \eta$, i.e., the normalized reflection coefficients of the spatial positions.

%------------------------------------------------------------------
\subsubsection{Reconstruction of Reflection Coefficients}
%------------------------------------------------------------------
Based on the compressive sensing technique described in Section~\ref{ssec: pce from rf signals}, we can obtain $\hat{\eta}$ by solving the following $l_1$-norm minimization problem:
\begin{align}
\text{(P2)}~
\min_{{\bm \eta}\in\mathbb C^{M}} \quad &\|{\bm \eta}\|_1,\\
& \|\bm H^*{\bm \eta} - \bm y_{\bm C^*} + \bm y^B_{\bm C^*}\|_2 \leq \epsilon. \nonumber
\end{align}
Here, $\epsilon$ is the variance of noise $\bm e$ in~(\ref{equ: practical compressive sensing}).
As~(P2) can be recast as a \emph{second-order cone program} problem, it can be solved using standard convex optimization tools in~\cite{Boyd_CONVEX}.
 
% The result vector $\hat{\bm \eta}$ is used to generate the point cloud.
From $\bm \eta$, the generation of the point cloud, which is a set of feature vectors and denoted by $\mathcal P$, is described as follows.
Set $\mathcal P$ contains $M$ elements, and each element is a $5$-dimensional feature vector, which is composed of its position and the reconstructed normalized reflection coefficient of it, i.e.,
\beq
\bm p_m = \left(
x_m, y_m, z_m, 
\mathrm{Re}({\bm \eta}_{m}),\mathrm{Im}({\bm \eta}_{m})
\right),~\forall m\in[1,M].
\eeq
Here, the first three dimensions of $\bm p_m$ indicate the coordinates of the $m$-th space block center, and $\mathrm{Re}$ and $\mathrm{Im}$ denote the real and imaginary parts of complex values, respectively.

%========================
\subsection{Semantic Segmentation}
%========================
\label{Segmentation}

%TODO !!! 这个section里面缺少对于training、loss function的描述。
 
\begin{figure}[!t] 
	\center{\includegraphics[width=0.6\linewidth]{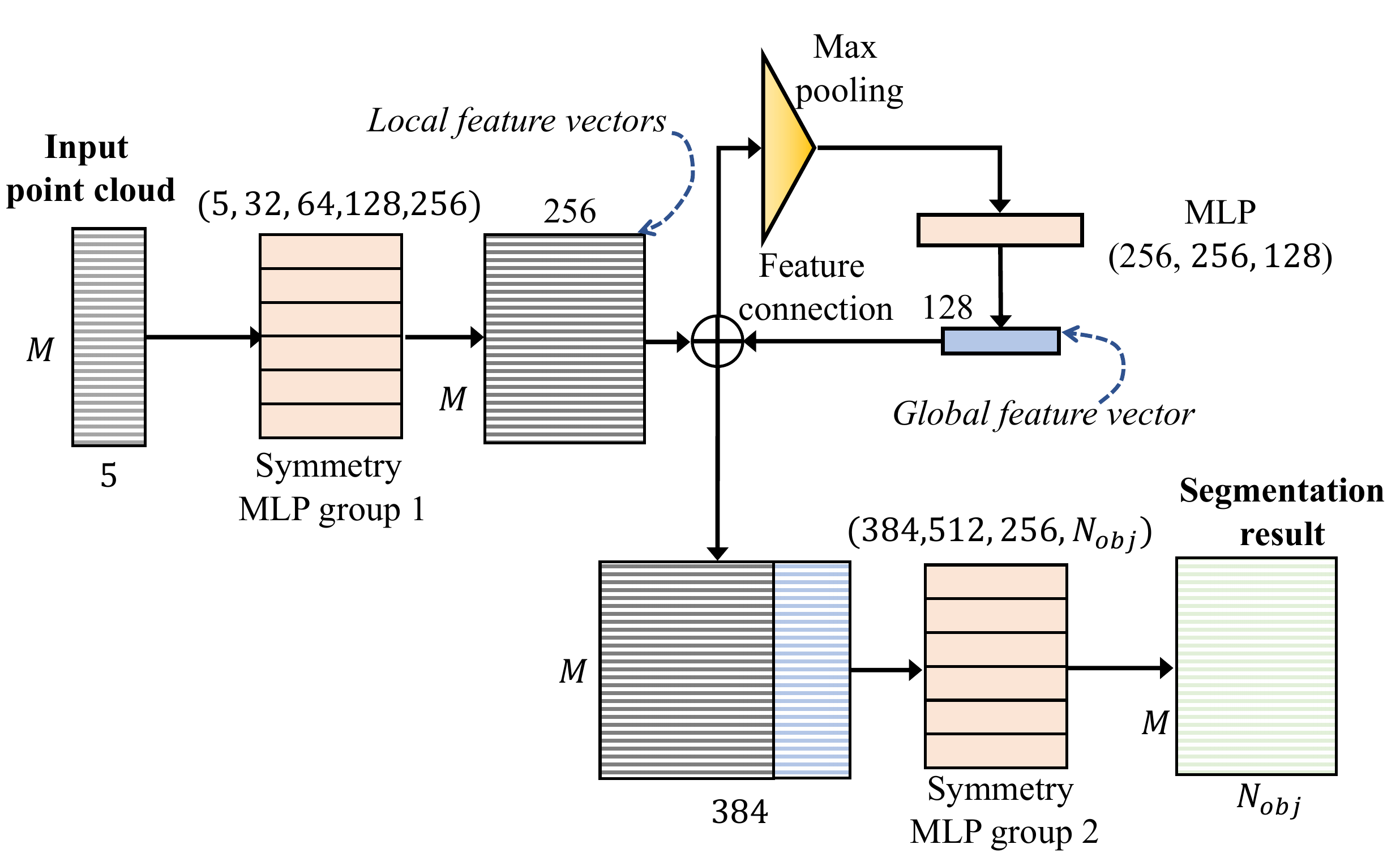}}
	\vspace{-1.5em}
	\caption{Workflow of semantic segmentation module.}
	\label{fig: seg. module}
\end{figure}

%In this section, we describe the semantic segmentation module of {\holosketch}.
% 简要介绍一下这个Module要做什么。
The semantic segmentation module takes the extracted point cloud $\mathcal P$ as input and outputs set $\tilde{\mathcal P}$, which contains the positions and semantic labels~(e.g., bottle, table, human, etc.) of the points in $\mathcal P$, i.e.,
\beq 
\tilde{\mathcal P} = \{(x_m, y_m, z_m, b_m)| m\in[1,M]\}
\eeq
where $b_m \in [1, N_{\obj}]$ denotes the semantic meaning of the $m$-th point, and $N_{\obj}$ denotes the total number of considered semantic meanings.
Without loss of generality, we denote the mapping performed by the semantic segmentation module by $\bm f_S: \mathcal P \rightarrow \hat{\mathcal P}$.

% 我们用什么样的方法来做
As described in Section~\ref{ssec: mlp}, $\bm f_S$ needs to be symmetric.
Besides, the process of labeling the semantic meaning of each point needs to consider both local and global information; because knowing the semantic meaning of the point cloud as a whole helps figure out the semantic meaning of each point.
Therefore, based on~\cite{qi2017pointnet}, we design the semantic segmentation module to contain \emph{symmetric MLP groups} and \emph{feature-gathering connections}, which are depicted in Fig.~\ref{fig: seg. module}.

%----------------------------------------------
\subsubsection{Symmetric MLP groups}  
%----------------------------------------------
We process $M$ points in $\mathcal P$ with $M$ symmetric MLPs with symmetric MLP groups.
As the MLP that treats each point has the same structure and adopts the same parameters, the results of the symmetric MLP group are invariant to the permutation of the input points.
Specifically, we adopt $2$ symmetric MLP groups in the semantic segmentation module.
The structures of the symmetric MLP groups are labeled in Fig.~\ref{fig: seg. module}.
%For example, $(5, 64, 64)$ beside the symmetric MLP group $1$ indicates that each MLP in this group contains $3$ layers, an input layer with size $5$ to receive the feature vector of a point~(containing its position and relative reflection coefficient), one hidden layer with size $64$, and an output layer with size $64$, that is, a new feature vector with $64$ dimensions is generated for the point.

%------------------------------------------------------
\subsubsection{Feature-gathering connections}
%------------------------------------------------------
Feature-gathering connections refer to the \emph{max pooling layer} and the concatenation of the local feature vectors and the global feature vector.
%TODO (Maybe) 可以试着修改一下程序，对global feature vector也可以得以训练。
%TODO !!! 这个max pooling到底是怎么实现的？它怎么就把很多的 point feature 转化为一个global feature了？这一点需要写清楚。
In the max-pooling layer, each dimension of the feature vectors of points are divided into small groups, and only the max values in the groups are picked as the output.
By this means, the max pooling layer reduces the amount of parameters and aggregate the information, which also alleviates overfitting. 
The output of the max-pooling layer is equal to the number of semantic classes in total and can be considered as the global feature vector.
Then, the global feature vector is concatenated to the local feature vector of each point.
By this means, the feature vector of each point now contains both local and global information.

%TODO !!! 这里是不是应该说一下loss方程，然后用一个最优化问题总结这个图啊。

%%%%%%%%%%%%%%%%%
\section{Implementation}
%%%%%%%%%%%%%%%%%
\label{sec: system implementation}

In this section, we present the implementation of {\holosketch}, including the implementation of the {\ris} and the RF transceiver module.
	
%xxxxxxxxxxxxx

%\begin{figure}[!t] % Example image
%	\center{\includegraphics[width=0.75\linewidth]{./figures/implementation/ris}}
%	\vspace{-1em}
%	\caption{{\ris} controller, {\ris}, and {\ris} element.}
%	\label{fig: RIS}
%\end{figure}

%\begin{table}[!t]
%
%\scriptsize
%%\vspace{-1em}
%\setlength{\belowcaptionskip}{-1.2em}
%\caption{{\ris} element's $S_{21}$ in different states.}
%\vspace{-1.2em}
%\centering
%\scriptsize
%\begin{tabular}{| c | c| c | c | p{1.3cm}<{\centering}| p{1.3cm}<{\centering} |}
%\Xhline{1.pt}
%\multirow{2}*{\textbf{State}} &
%\multicolumn{3}{c|}{\textbf{Bias Voltages}}  &
%\multicolumn{2}{c|}{\textbf{$S_{21}$}}\\ 
%\cline{2-6} 
%	& \textbf{PIN \# 1} &\textbf{PIN \#2} &\textbf{PIN \#3} & \textbf{Phase} & \textbf{Amplitude}\\
%\hline
%$\hat{s}_1$ & $0$V& $0$V& $0$V & $\pi/4$ & $0.97$ \\
%$\hat{s}_2$ & $0$V& $0$V& $1.2$V & $3\pi/4$ & $0.97$ \\
%$\hat{s}_3$ & $1.2$V& $0$V& $1.2$V & $5\pi/4$ & $0.92$ \\
%$\hat{s}_4$ & $1.2$V& $1.2$V& $0$V& $7\pi/4$ & $0.88$ \\
%\Xhline{1.pt}
%\end{tabular}
%\label{table: S21 at 4 states}
%\end{table}
%xxxxxxxxx

\begin{figure}[!t]
	\center{\includegraphics[width=0.6\linewidth]{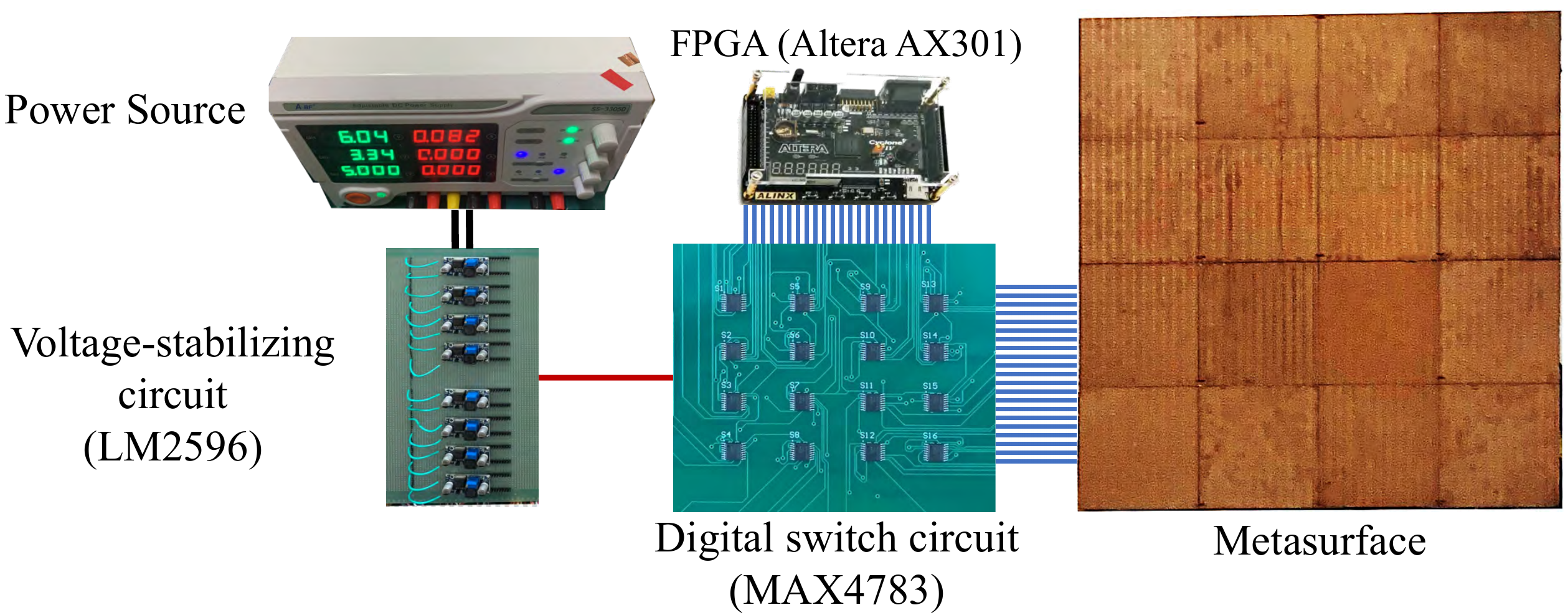}}
	\vspace{-1.em}
	\caption{Diagram of {\ris} control circuit.}
	\label{fig: RIS control circuit}
\end{figure}

%=================================
\subsection{Building the {\Ris}}
%=================================
%We adopt the electrically modulated {\ris} in~\cite{Li2019Machine}.
%TODO 要加上Li2019Machine这个文章的引用
As shown in Fig.~\ref{fig: RIS control circuit}, the {\ris} is with the size of $69\times 69\times 0.52$ cm$^3$ and is composed of $16$ independently controllable groups which are tightly paved in squares.
Each group contains $12*12 = 144$ {\ris} elements arranged in a two-dimensional array, and thus the total number of {\ris} elements is $2304$.

Each {\ris} element has the size of $1.5\times 1.5 \times 0.52$ cm$^3$ and is composed of $4$ rectangle copper patches printed on a dielectric substrate~(Rogers 3010) with a dielectric constant of $10.2$ and $3$ PIN diodes~(BAR 65-02L).
Any two adjacent copper patches are connected by a PIN diode, and each PIN diode has two operation states, i.e., ON and OFF, which are controlled by applied bias voltages on the via holes.
When the applied bias voltage is $1.2$ V~(or $0$ V), the PIN diode is at the ON~(or OFF) state.
% 以下这两句话太过于详细，其实可以不加。对于文章也并不重要。
%Besides, to isolate the DC feeding port and microwave signal, four choke inductors of $30$ nH are used in each {\ris} element. 
%Besides, as shown in Fig.~\ref{fig: RIS}, a {\ris} element contains four choke inductors that are used to isolate the DC feeding port and RF signals.

As there are $3$ PIN diodes in a {\ris} element, the total number of possible states of a {\ris} element is $8$.
We simulate the $S_{21}$ parameters, i.e., the \emph{forward transmission gain}, of the {\ris} element in different states for normal-direction incident RF signals in CST software, Microwave Studio, Transient Simulation Package~\cite{Hirtenfelder2007Effective}. 
%Table~\ref{table: S21 at 4 states} provides the amplitudes and angles of $S_{21}$ of a {\ris} element in $4$ of $8$ states for incident sinusoidal signals with frequency $3.2$~GHz.
We pick these four states with a phase shift interval equaling to $\pi/2$ as the \emph{available state set} $\mathcal S_a$, i.e., $\mathcal S_a = \{\hat{s}_1,\hat{s}_2,\hat{s}_3,\hat{s}_4\}$.
The four selected states have the phase values equaling to $\pi/4$, $3\pi/4$, $5\pi/4$ and $7\pi/4$, respectively.
%TODO 在附录写的时候，就要体现出 available state set 这一个名词

As described in Section~\ref{ssec: ris model}, the {\ris} elements within the same group are in the same state.
The states of the $16$ groups are controlled by the {\ris} control circuit, which contains a direct current power supply, multiple voltage-stabilizing modules~(LM2596), $16$ digital switch circuit~(MAX4783), and a field-programmable gate array~(FPGA)~(ALTERA AX301).
%这些元件是怎么控制{\ris}的。
The DC power supply is connected to the voltage-stabilizing modules, and the input voltage to the voltage-stabilizing modules is about $6$~V.
The voltage-stabilizing modules stabilize the input voltage and reduce it to a $1.2$ V output level. 
The digital switch circuits are single-pole double-throw and control whether the PIN diodes of {\ris} elements are connecting to the ground, i.e., $0$ V level, or to the $1.2$ V output of the voltage-stabilizing modules.

\begin{figure}[!t] % Example image
	\center{\includegraphics[width=0.55\linewidth]{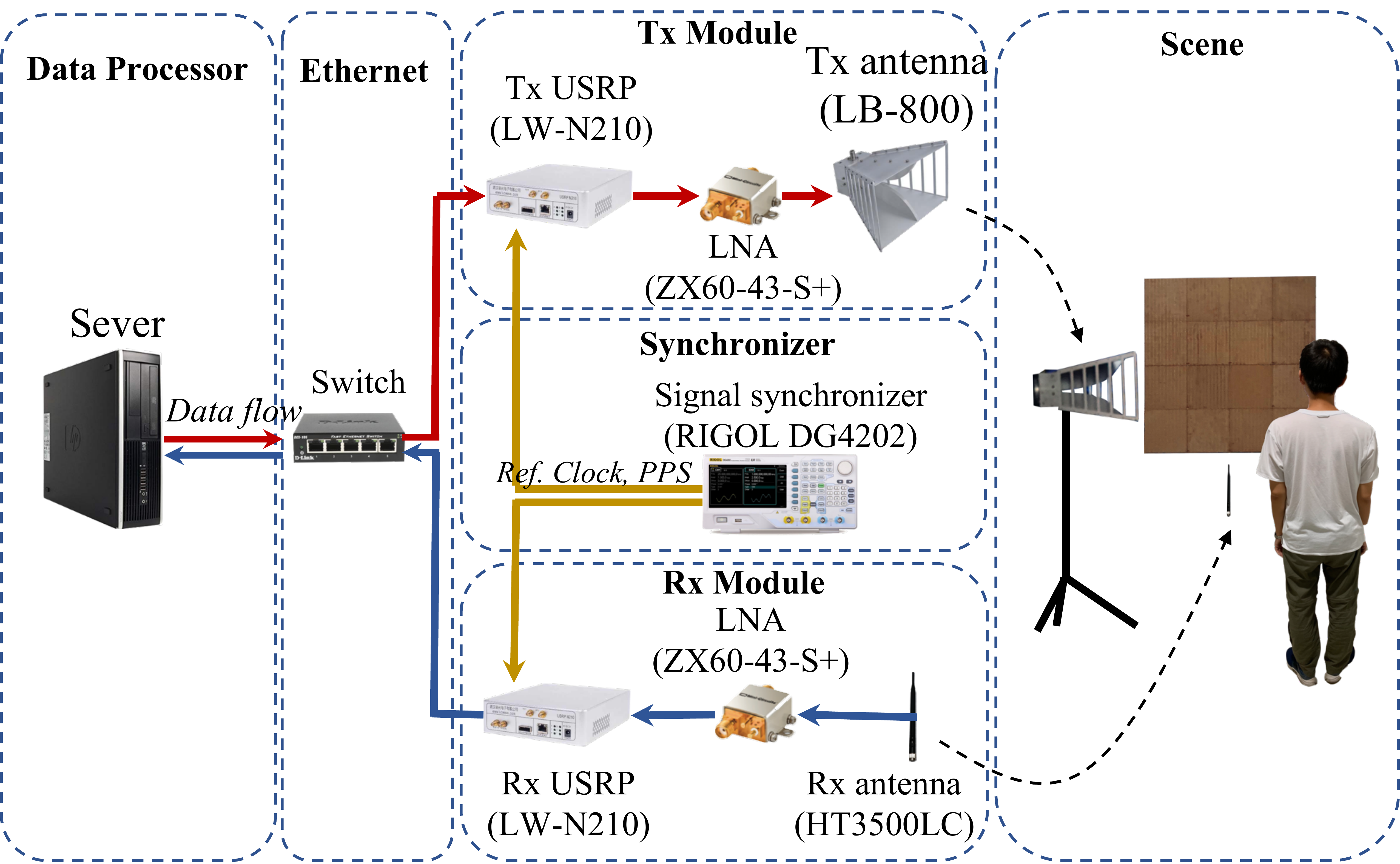}}
	\vspace{-1.em}
	\caption{Components of the transceiver module.}
	\label{fig: transceiver}
	%TODO 线加粗，不要画LNA，可以放上实景照片代替超平面和人体的部分。
	%TODO 这个图一定要变化一下，目前和JSAC实在是一样，这样不行。
\end{figure}

%==========================================
\subsection{Building the Transceiver Module}
%==========================================
As shown in Fig.~\ref{fig: transceiver}, we build the RF transceiver module in {\holosketch} by using the following components.

\parhead{(1) USRP devices}: We implement the Tx and Rx based on two~USRPs~(LW-N210), which are capable of converting baseband signals to RF signals, and vise versa.
The USRP is composed of the hardwares, including the RF modulation/demodulation circuits and baseband processing units, and can be controlled by using the GNU packet in Python~\cite{blossom2004gnu}.

\parhead{(2) Low-noise amplifiers~(LNAs)}: Since the RF signals need to be reflected twice~(on {\ris} and on objects) before reaching the Rx antenna, they suffer from large attenuation in signal strength, which results in low SNR and degrades the measurement accuracy.
To handle this issue, two LNA~(ZX60-43-S+) connect the Tx/Rx USRP and the Tx/Rx antenna and amplify the transmitted/received RF signals by about $15$ dB.

\parhead{(3) Tx and Rx antennas}: The Tx antenna is a directional double-ridged horn antenna~(LB-800), and the Rx antenna an omni-directional vertical antenna~(HT3500LC). 
The polarization of both the Tx and Rx antennas is linear and vertical to the ground.

\parhead{(4) Signal synchronizer}: For the Rx USRP to obtain the relative phases and amplitudes of the received signals with respect to the transmitted signals of the Tx USRP, we employ a signal source~(DG4202) to synchronize the frequency and phase of the Tx and Rx USRPs. The signal source provides the reference clock signal and the pulses-per-second~(PPS) signal to the USRPs, which ensures the modulation and demodulation of the USRPs to be coherent. 

\parhead{(5) Ethernet switch}: The Ethernet switch connects the USRPs and a sever forming a local Ethernet, where the controlling signals and received signals are exchanged.

\parhead{(6) Sever}: The sever controls the two USRPs by using the GNU packet in Python, extracts the measurement vectors from the received signals of the Rx USRP, and performs point cloud extraction and semantic segmentation.

%\end{itemize}
%TODO (Maybe) 是不是可以写一下path0的链路？

%%%%%%%%%%
\section{Simulation and Experimental Evaluation}
%%%%%%%%%%
\label{sec: evaluation}
In this section, we demonstrate the experimental setup for {\holosketch} and evaluate the performance of the three component modules.

\begin{figure}[!t] % Example image
	\center{\includegraphics[width=0.45\linewidth]{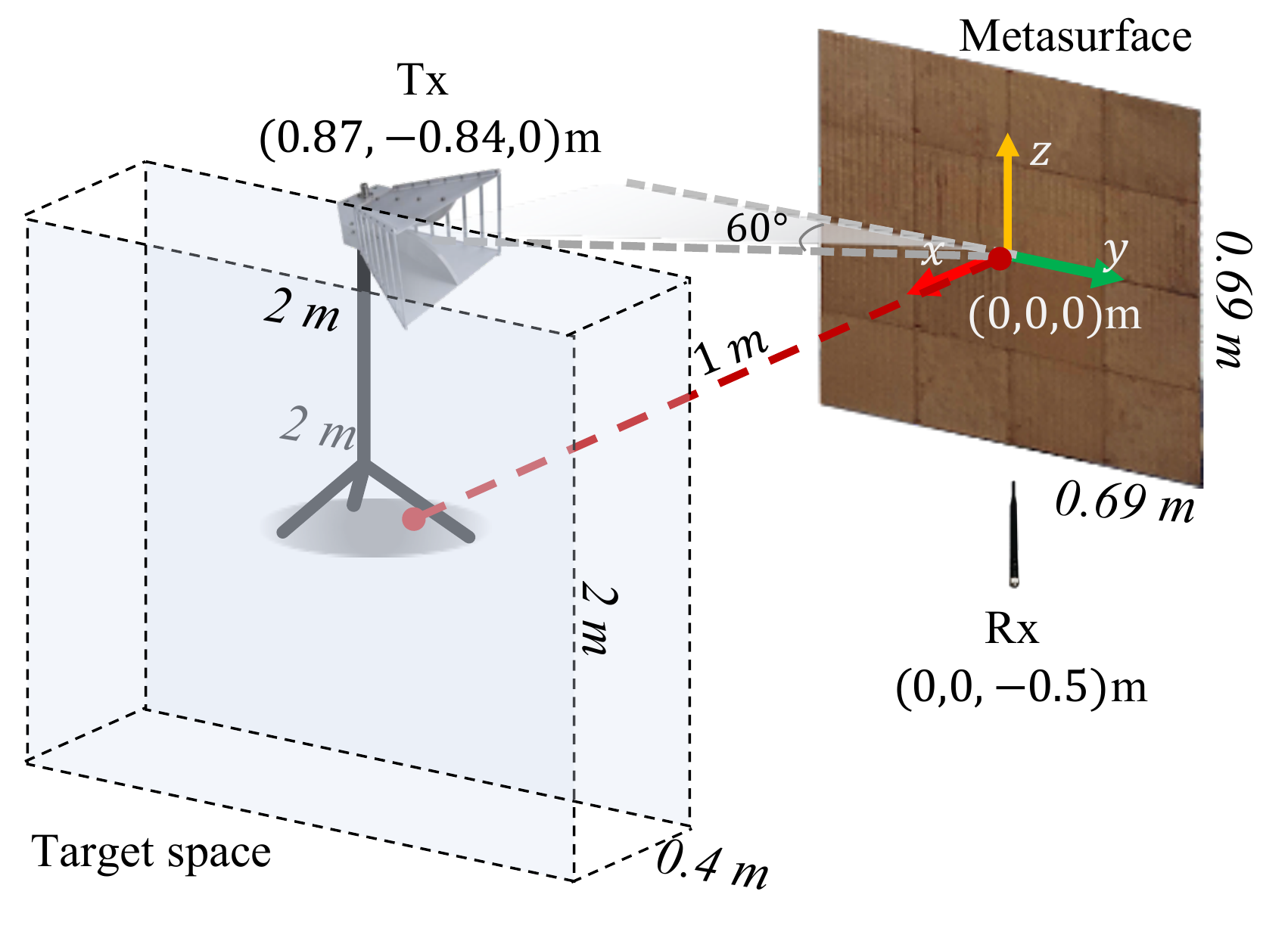}}
	\vspace{-1em}
	\setlength{\belowcaptionskip}{-1.2em}   %调整图片标题与下文距离
	\caption{Environment layout in experiments.}
		\vspace{-1em}
	\label{fig: environment setup}
\end{figure}

%======================
\subsection{Experimental Setup}
%======================
We describe the experimental setup in three aspects: the environment layout to test {\holosketch}, the data structure of each module, and the adopted evaluation metrics. 

%------------------------------------------
\subsubsection{Environment layout}
%------------------------------------------
%TODO 这里主要是叙述一下环境布置，应该加一张环境布置的图来说明比较好。如果地方允许的话。
The environment layout of {\holosketch} is shown in Fig.~\ref{fig: environment setup}.
To be specific, the origin of coordinate is at the center of the {\ris}, and the {\ris} is in the $y-z$ plane.
Besides, the $z$-axis is vertical to the ground and pointing upwards, and the $x$- and $y$-axes are parallel to the ground.
The Tx and Rx antennas are located at $(0.87, -0.84, 0)$~m and $(0,0,-0.5)$~m, respectively.

The humans and objects are in the space of interest, which is a cuboid region located at $1$~m from the {\ris}.
Since the space of interest is behind the Tx antenna and the Tx antenna is directional horn antenna, no LoS signal path from the Tx antenna to the space of interest exists.

The humans and objects for recognition are located within a \emph{target space}, which is a $0.4\times 2\times 2$ m$^3$ cuboid space.
The target space is regularly divided into $M=400$ space blocks each with size $0.4\times 0.2\times 0.2$ m$^3$.

%-----------------------------------
\subsubsection{Collected Data}
%-----------------------------------
The optimized configuration matrix of {\ris}, i.e., $\bm C^*$, is obtained by solving~(P1) in the sever and is uploaded to the FPGA.
In the data collection phase, the {\ris} changes its configuration by $0.1$ second. 
To obtain the corresponding measurement matrix, i.e., $\bm H^*$, we set a metal patch with size $0.2\times 0.2$ m$^2$ at the center of each space block sequentially given {\ris} using $\bm C^*$, as described in Section~\ref{sec: obtain H}.

We first generate a set of $64$ point clouds with semantic labels as the ground truth set.
Then, we arrange humans and objects in the target space according to each of the point clouds.
The objects include a bottle, a laptop, and a suitcase.
We measure the received signals following the protocol in Section~\ref{ssec: operating protocol}.
Using measurement matrix $\bm H^*$ corresponding to $\bm C^*$, point cloud extraction module processes the received signals by solving~(P2). 
The ground truth set and the corresponding extracted point clouds constitute the training data for the semantic segmentation module.
 
In the collected training data, each point is represented by a $5$-dim vector and a label.
The first three dimensions indicate the coordinate of the point; the next $2$ dimensions indicate the real and imaginary values of the regenerated reflection coefficients of the point.
The label takes value in set $[1,N_{\obj}]$ where $N_{\obj} = 5$, which represent human, bottle, laptop, suitcase, and empty space, respectively.

%----------------------------------------
\subsubsection{Evaluation metrics}
%----------------------------------------
We adopt the following three evaluation metrics.
%\begin{itemize}
\indent (a) \emph{AMC}: As defined in~(\ref{equ: mutual coherence}), the AMC evaluates the average coherence between every two columns in the designed measurement matrix.
A lower AMC indicates the propagation channels via different space blocks are more independent of each other.
The AMC is inversely proportional to the reconstruction performance of the compressive sensing method~\cite{elad2007optimized}.
\indent (b) \emph{Loss}: We adopt \emph{cross-entropy loss}~\cite{goodfellow2016deep} as the metric to train the semantic segmentation module. 
The cross entropy loss is defined as $\mathcal L_{\bm b'}(\bm b) = \sum_i^{N_{\obj}} b_i'\log(b_i)$, where $\bm b'$ is a $0-1$ vector indicating the true label of a point, and $\bm b$ is the probability vector obtained by the semantic segmentation module. 
\indent (c) \emph{Average error rate}: For each label, the error rate is defined as the ratio between the number of inaccurately labeled points and the total number of points with that label in truth. 
We adopt the average of the error rates of the $N_{\obj}$ labels as the metric to evaluate the performance of {\holosketch}.
%是否要相应的修改training target?

%\end{itemize}

%============================================
\subsection{AMC by Radio Environment Reconfiguration}
%============================================

\begin{figure}[!t] 
	\center{\includegraphics[width=0.45\linewidth]{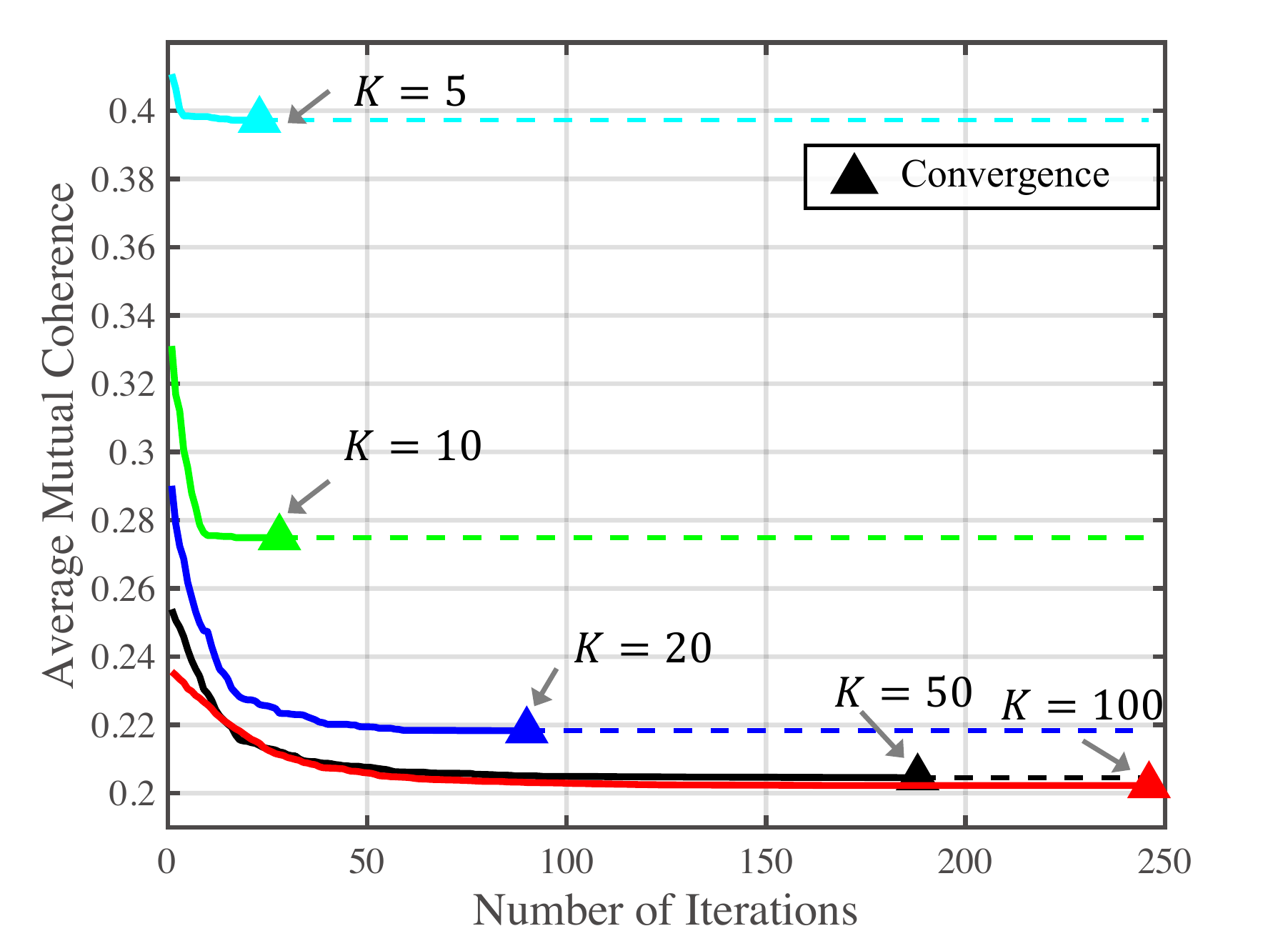}}
	\vspace{-1.em}
	\setlength{\belowcaptionskip}{-1.2em}   %调整图片标题与下文距离
	\caption{AMC of measurement matrix vs. the number of training iterations.}
	\label{fig: cohereVSitrnum}
	%TODO 倒数第二根线要改成K=50
\end{figure}

Fig.~\ref{fig: cohereVSitrnum} shows the AMC of $\bm H$ vs. the number of iterations in Algorithm~1, under different values of the number of configurations, $K$.
It can be observed that AMC decreases with the number of iterations, which verifies the effectiveness of the proposed radio environment reconfiguration algorithm.
Besides, it can also be seen that the converged optimal AMC of $\bm H$ decreases with $K$.
This can be explained as follows.
To reduce the mutual coherence of $\bm H$, it requires different columns of $\bm H$ have large elements at different dimensions.
As $K$ is the dimension of $\bm h_m$, large $K$ increases the probability to have the large elements at different dimensions and reduce the mutual coherence.
Therefore, as $K$ increases, the AMC value of the optimized measurement matrix decreases, which can lead to a higher accuracy for the compressive sensing technique to extract exact point clouds.

\begin{figure}[!t] 
	\center{\includegraphics[width=0.55\linewidth]{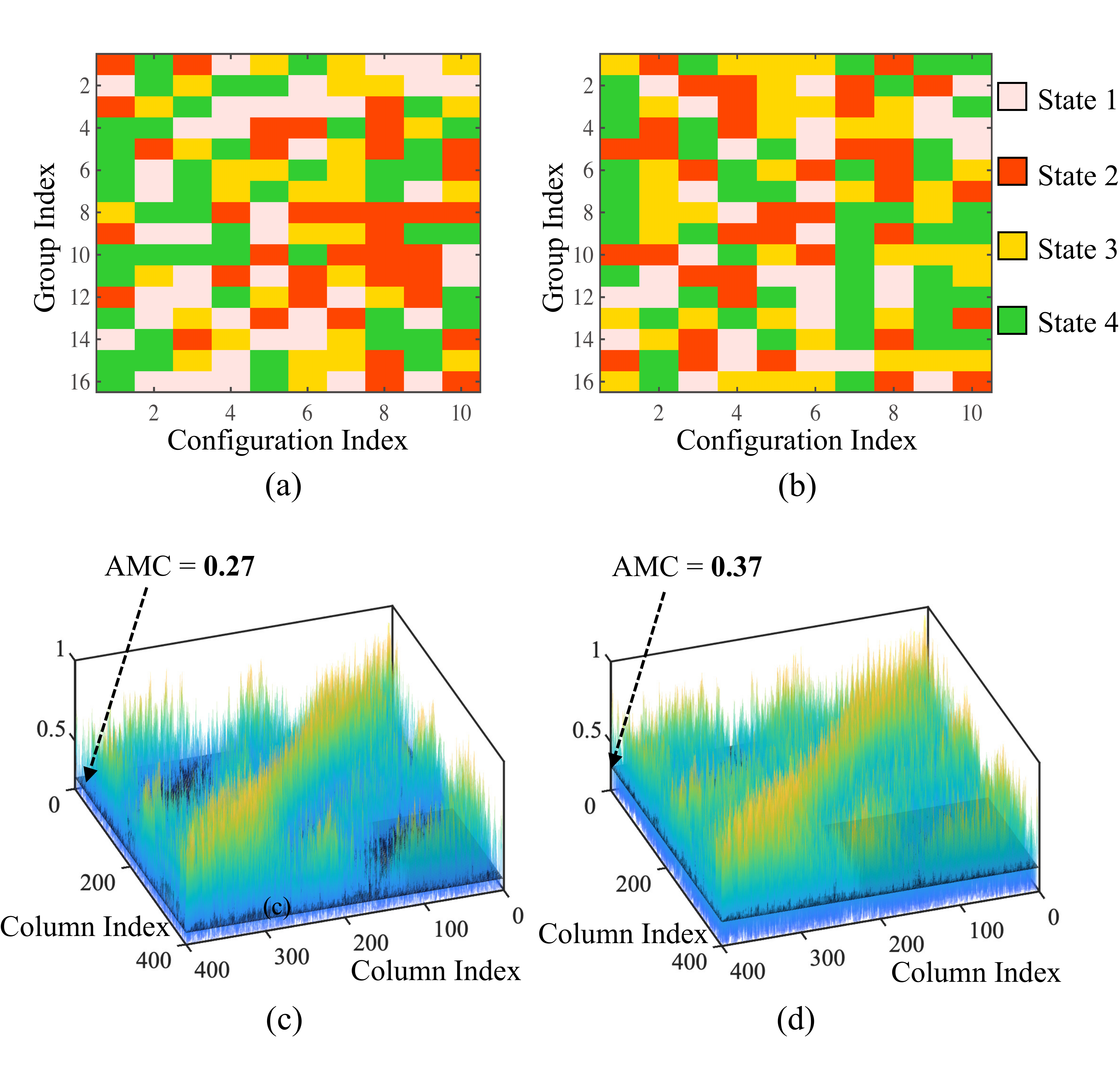}}
	\vspace{-1em}
	\setlength{\belowcaptionskip}{-1.2em}   %调整图片标题与下文距离
	\caption{Illustrations on the (a)~optimized and~(b) random {\ris} configuration matrices, and (c)~the mutual coherence values of the optimized configuration matrix and (d)~those of the random configuration matrix. The planes in~(c) and~(d) indicate the AMC values.}
	\label{fig: config sequence compare}
	%TODO 把这个图的次序改一下，并且这个图的画法其实不是很清楚。
\end{figure}

We compare the mutual coherence of the measurement matrices corresponding to the random and optimized configuration matrices.
The configuration matrix in Fig.~\ref{fig: config sequence compare}~(a) is $\bm C^*$ obtained by using Algorithm~1, and Fig.~\ref{fig: config sequence compare}~(c) shows the mutual coherence of corresponding $\bm H^*$.
Besides, the configuration matrix in Fig.~\ref{fig: config sequence compare}~(a) is a random configuration matrix where the elements in $\bm C$ are generated following uniform distribution in $[1,4]$.
Fig.~\ref{fig: config sequence compare}~(d) shows the coherence of column vectors of $\bm H$ corresponding to random $\bm C$ in Fig.~\ref{fig: config sequence compare}~(a).
Comparing Figs.~\ref{fig: config sequence compare}~(c) and~(d), we can observe that Algorithm~1 optimizes the configuration matrix, which effectively reduces the mutual coherence of the measurement matrix, resulting in a lower AMC than that of a random configuration matrix.
Based on discussion in Section~\ref{RadioReconfiguration}, the configuration matrix in Fig.~\ref{fig: config sequence compare}~(a) can result in higher accuracy of point cloud extraction than that in Fig.~\ref{fig: config sequence compare}~(b).

%===================================
\subsection{Extracted Point Clouds}
%===================================
\begin{figure}[!t] 
	\center{\includegraphics[width=0.75\linewidth]{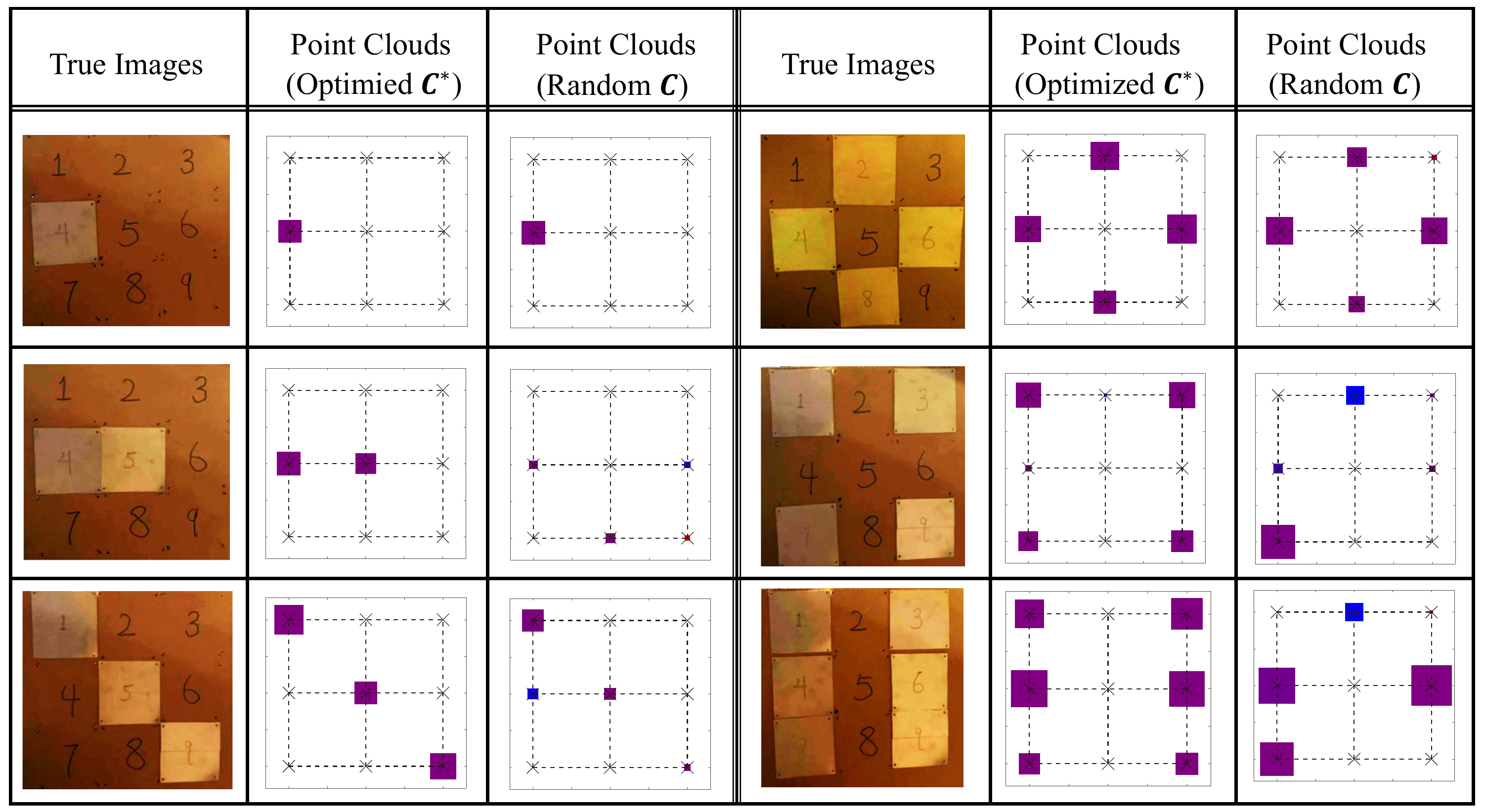}}
	\vspace{-1.em}
	\caption{Extracted point clouds by using random and optimized configuration matrices. A large square indicates a large absolute value of the obtained reflection coefficient at a certain position, and the colors of squares indicate the phases.}
			\setlength{\belowcaptionskip}{-.0em}   %调整图片标题与下文距离
	%TODO true image改名叫做photo
	\label{fig: pc with rand and opt config}
\end{figure}

Fig.~\ref{fig: pc with rand and opt config} shows the photos and the corresponding extracted point clouds under random and optimized configuration matrices.
%一下浅色的地方是金属片，而深色、棕色的地方是纸板，它们都放在物体平面上。
In the photos, the light~(yellow) regions are the metal patches, and the dark~(brown) regions are the cardboards which have a negligible impact on wireless signals.
% 在空间中金属片点数少的时候
It can be observed that when the number of metal patches is small~(the number of metal patches is less than $3$), the point cloud extraction module can successfully reconstruct the point clouds which reflect the true images well.
However, when the number of metal patches is larger than $4$, the point clouds are not in accordance with true images.
This is due to that the compressive sensing method requires the target vector to be sparse.
Besides, the point clouds obtained by using the optimal {\ris} configuration $\bm C^*$ reflects the photos more accurately than those obtained by using a random {\ris} configuration $\bm C$, which verifies the effectiveness of optimizing radio environment reconfiguration.

%===================================
\subsection{Semantic Segmentation of Human and Objects}
%===================================

\begin{figure}[!t] \center{\includegraphics[width=0.7\linewidth]{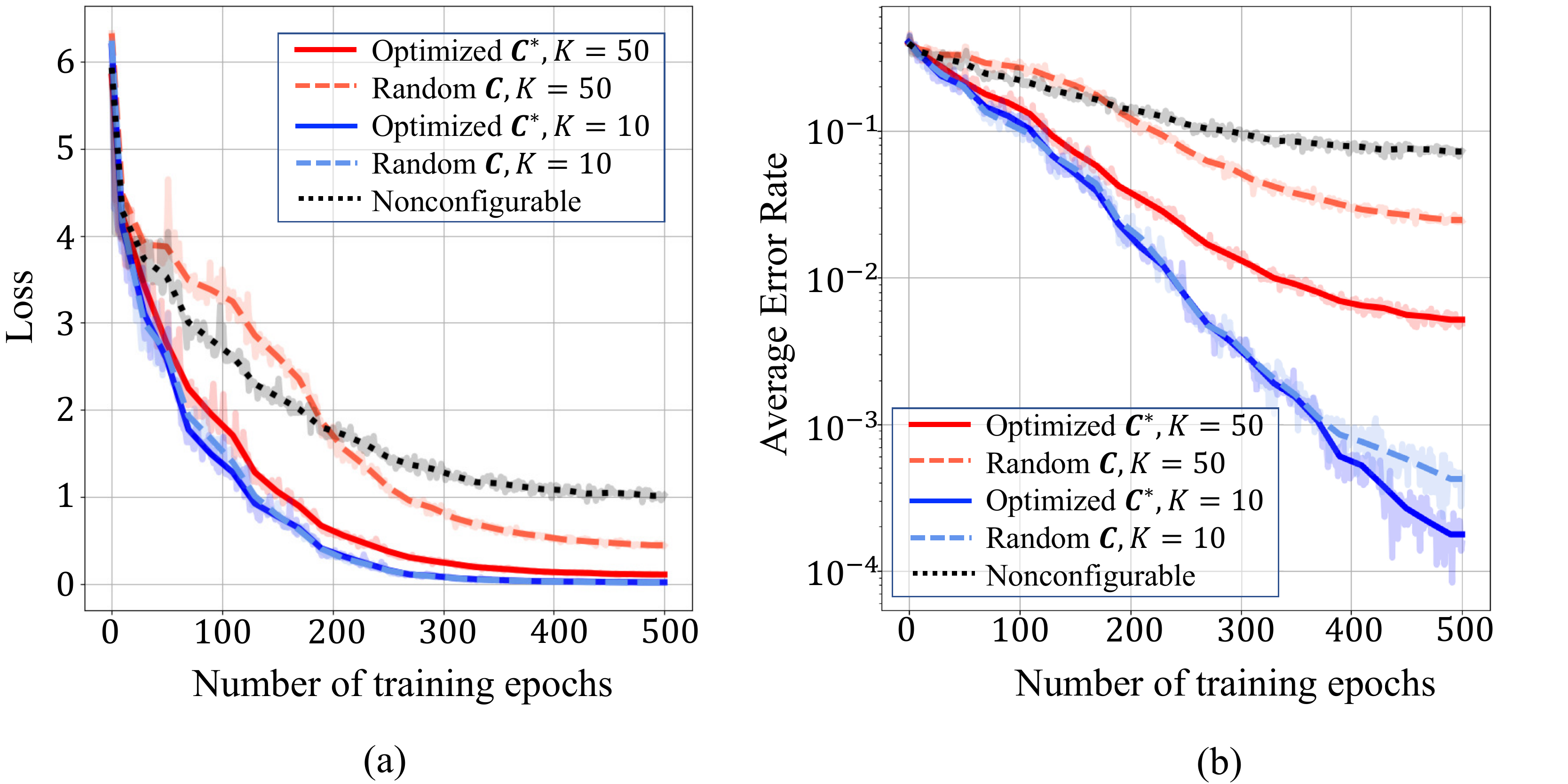}}
	\vspace{-1.5em}
	\caption{
	(a) Loss and~(b) accuracy vs. the number of training epochs with optimized and random configurations under different $K$. The shaded areas denote the values for each epoch, and the thick lines denote the average values for the adjacent $20$ epochs.}
	\label{fig: segmentation accuracy}
	%TODO 要根据这句图相应修改说明
\end{figure}

Figs.~\ref{fig: segmentation accuracy}~(a) and~(b) shows the loss and average error rate vs. the number of training epochs of the semantic segmentation module.
The solid and dash lines in red and in blue are the loss and average error rate curves obtained when $K=10$ and $K=50$, respectively, where solid lines are the results under $\bm C^*$ and dash lines are the results under $\bm C$.
The dot lines in black indicate those curves obtained in a nonconfigurable radio environment, where the configuration of the {\ris} is fixed to all state $\hat{s}_1$, i.e., $K=1$ and $\bm C = \bm 1$.
As the number of training epochs increases, both the loss and the average error rate decreases.
Compared to those in a nonconfigurable radio environment where the configuration of the {\ris} is fixed, the loss and average error rates in configurable radio environments are significantly lower, which verifies the effectiveness of using a {\ris} to configure radio environment.
Besides, when $K$ is small~($K=10$), it can be observed that using $\bm C^*$ can help the semantic segmentation module to train the MLP with a much lower loss and error rate.
When $K=10$, after about $350$ epochs of training, {\holosketch} can perform semantic segmentation with an average error rate of less than $1\%$.
If the number of configurations is sufficiently large, i.e., $K=50$, the average error rate can be further reduced to less than $0.1\%$ after $400$ epochs of training.
However, when $K=50$, the data collection phase lasts for $5$~seconds, which makes the assumption of humans and objects being static during the data collection phase impractical.
%TODO 要说明一下使用的loss function,以及训练使用的激活方程等等。
%TODO 简单提一下我们使用了什么样的学习率下降和优化函数和激活函数。

%Fig.~\ref{fig: segmentation accuracy} shows the mutual coherence of $\bm H$ and segmentation accuracy versus the number of configurations when random and optimized configuration matrix $\bm H$ are used.
%It can be observed that with the increment of the number of configurations $K$, the mutual coherence of $\bm H$ decreases and the segmentation accuracy increases.
%It shows in both random $\bm H$ and optimized $\bm H$ cases that increasing the number of configurations can lower the mutual coherence of $\bm H$, and improve the segmentation accuracy.
%Moreover, by comparing the results of the random $\bm H$ and that of the optimized $\bm H$, it shows that minimizing the mutual coherence of $\bm H$ by the proposed radio environment reconfiguration algorithm can significantly improve the segmentation accuracy, especially when $K$ is large. 

\begin{figure}[!t] 
	\center{\includegraphics[width=0.5\linewidth]{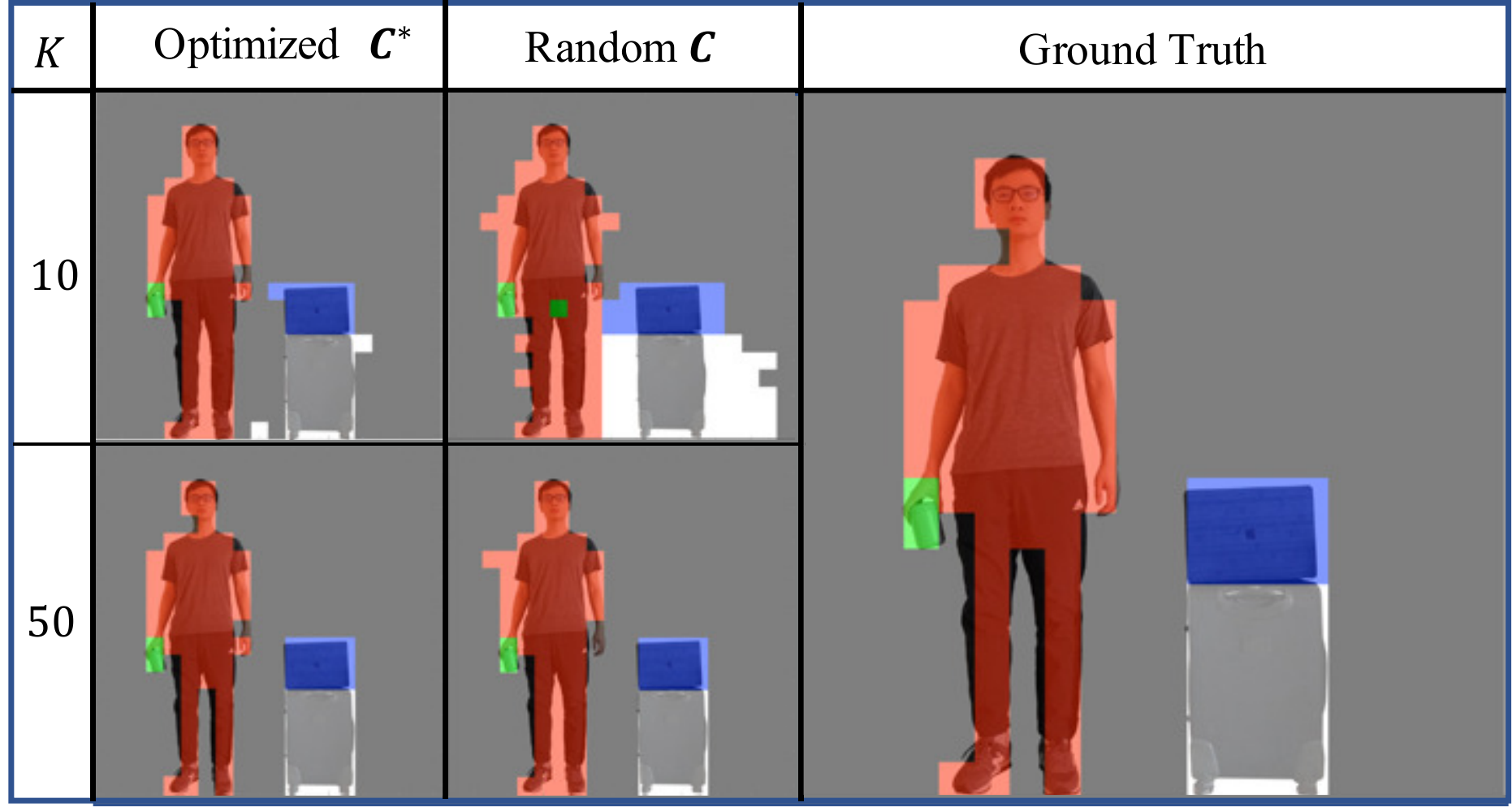}}
	\vspace{-0.5em}
	\caption{Semantic segmentation results for human and objects given that the {\ris} adopts optimized and random configuration matrices with different $K$. The number of training epochs is $500$. The human, suitcase, laptop, and bottle are labeled by red, white, blue, and green colors, respectively.}
	\label{fig: segmentation results}
\end{figure}

%TODO 这里为了匿名性我们是不是要用背部的图片呢。
\begin{figure*}[!t] 
%	\vspace{-2.em}
	\center{\includegraphics[width=0.92\linewidth]{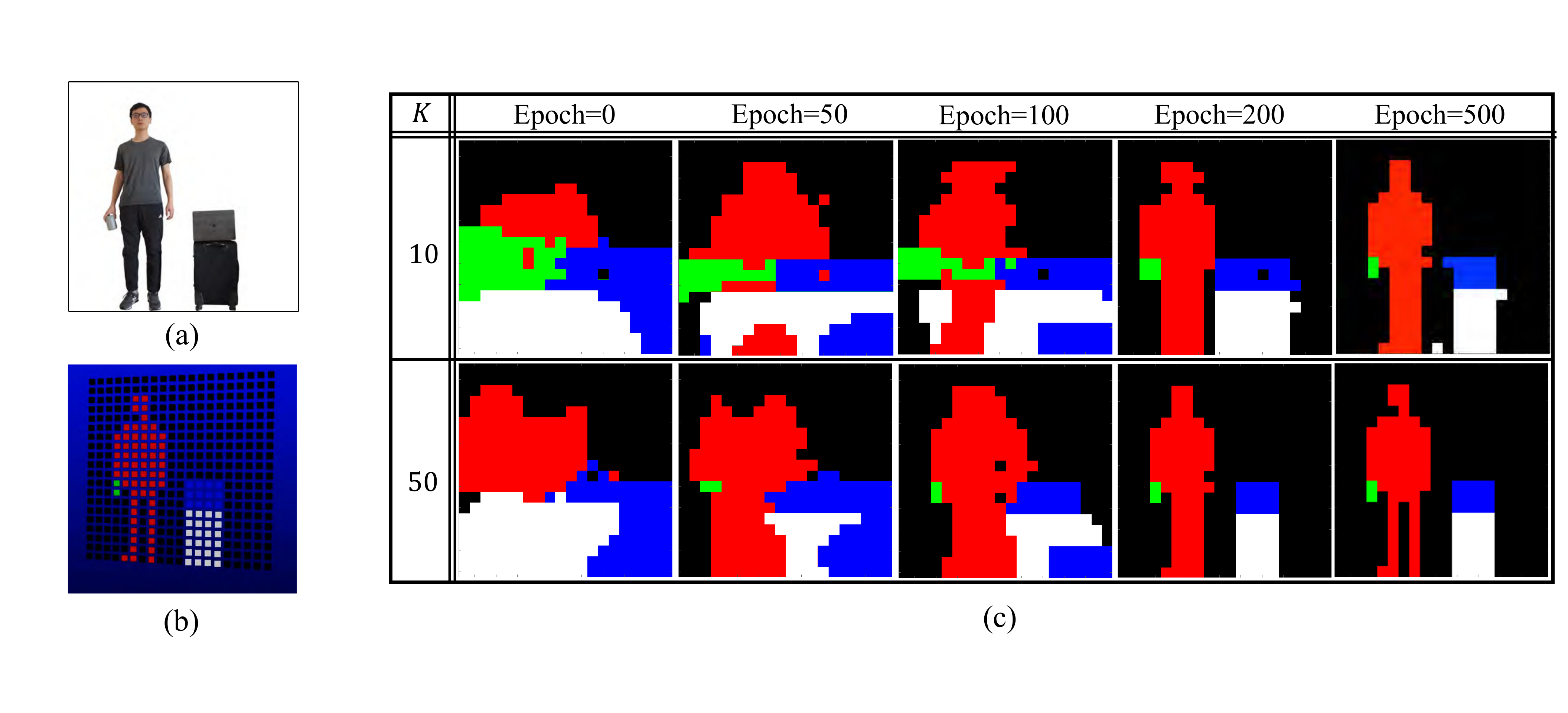}}
	\vspace{-2.4em}
	\caption{(a) and (b) are the image of the human and objects and the ground truth of the corresponding labeled point cloud, respectively, and (c) is the semantic segmentation results in different training epochs given {\ris} adopting the optimized $\bm C^*$ with $K=10$ and $K=50$.}
	\label{fig: pc obj human}
	%TODO 这个图可以去掉一行
\end{figure*}

Fig.~\ref{fig: segmentation results} shows the semantic segmentation results after training for $500$ epochs overlaid on the ground truth images, given that {\ris} adopts optimized and random configuration matrices with different $K$.
Moreover, Fig.~\ref{fig: pc obj human} provides the details of the training process by showing the semantic segmentation results in different training epochs.
Figs.~\ref{fig: pc obj human}~(a) and~(b) are the images of the human and objects, and the ground truth of the corresponding labeled point cloud, respectively, and~(c) shows the semantic segmentation results in different training epochs, where the {\ris} adopts $\bm C^*$ with $K=10$ and $K=50$.
In the semantic segmentation results, the human, suitcase, laptop, and bottle in the results are labeled by red, white, blue, and green colors, respectively.

In Fig.~\ref{fig: segmentation results}, comparing the cases where $K=50$ and $K=10$, we can observe that increasing the number of configurations in a cycle enables the {\holosketch} to obtain labeled point cloud closer to the ground truth.
In Fig.~\ref{fig: pc obj human}~(c), it can be seen that when $K$ is larger, the training process is faster, as the semantic segmentation result gets close to the ground truth in an earlier epoch.
Besides, in Fig.~\ref{fig: segmentation results}, it can also be seen that using the optimized configuration matrix improves segmentation accuracy.
When $K=50$, the improvement due to using optimized configuration matrix is smaller than when $K=10$. 
Nevertheless, since large $K$ results in a long duration of the data collection phase and thus low recognition speed of {\holosketch}, a small $K$ is preferable, where adopting an optimized configuration matrix is necessary and important.

%\vspace{-1em}
%%%%%%%%%%
\section{Conclusion}
%%%%%%%%%%
\label{sec: conclu}
In this paper, we have presented {\holosketch}, a {\ris}-based RF-sensing system, to perform semantic segmentation for humans and objects in 3D space.
We have designed the {\holosketch} with three modules, i.e., a radio environment reconfiguration module, a point cloud extraction module, and a semantic segmentation module.
{\holosketch} can actively modify the radio environment according to the configurations of a {\ris} and generate abundant favorable propagation channels for sensing, which have been optimized by using the proposed configuration optimization algorithm.
By the point cloud extraction module, {\holosketch} extracts the point cloud in space, which can be further processed by its semantic segmentation module for semantic meaning labeling.

%从实验效果说
%1. 我们设计的算法降低了选出的信道的自相关，从而提升了点集提取的准确度。
Our results have shown that, firstly, the {\ris}-based radio environment reconfiguration module with the proposed algorithm can produce measurement matrices with low AMC, which promotes the accuracy of point cloud extraction.
%2. {\holosketch}可以有效的对空间点进行label，正确率可以达到90以上。
Secondly, after training, {\holosketch} can label semantic meanings of the points with an average error rate of less than $1\%$, given the setup of a human, a suitcase, a laptop, and a bottle in a $1.6~m^3$ indoor space represented by $400$ points.
%3. 采用经过优化的H，可以降低训练所需的次数，或者需求的测量次数。
Thirdly, optimizing the measurement matrix has reduced the required number of training epochs and measurements to obtain an accurate segmentation.

% Disscussion Part
For further making {\holosketch} system complete, it requires the following additional technical challenges to be addressed.
\begin{itemize}[leftmargin=*]
\item \textbf{Higher resolution of {\holosketch}}:
In this paper, the target space is divided into $0.4\times 0.2\times 0.2$ m$^3$ cubic regions, i.e., space blocks, which limits the resolution.
Increasing the resolution of {\holosketch} requires the target space to be finely divided.
As described in Section~\ref{sec: obtain H}, to obtain the measurement matrix, a metal patch needs to be placed in each space block.
Therefore, to increase the resolution of {\holosketch} implies the size of the metal patch needs to be shrunk and the number of measurements needs to be increased.
% 此外，小金属片使接收到的反射信号较小，因此需要更高增益的天线或者LNA。（不过这个上面说过了这里再说有点重复）=
\item \textbf{Faster data collection of {\holosketch}}:
In the data collection phase, the {\ris} changes its configuration by $0.1$ second, and thus the data collection phase lasts for at least $K/10$ seconds.
As the humans and objects need to be static during the data collection phase, a long data collection phase due to large $K$ requires more efforts of ensuring them stay still when deploying {\holosketch}.
To speed up the data collection procedure, we can adopt more advanced switching circuits and FPGA to shorten the time for changing the configurations of {\ris}.
\end{itemize}

 % APPENDIX
 \begin{appendices}
 \section{Calculation of Measurement Matrix}

Given configuration matrix $\bm C$, we now calculate the corresponding measurement matrix $\bm H$.
Based on ray-tracing technique~\cite{goldsmith2005wireless} and~(\ref{equ: math of RIS}), we first calculate channel gain matrix $\bm A$, where the elements indicate the channel gains of the radio paths from the Tx to Rx via the $L$ {\ris} groups in $N_s$ states and the $M$space blocks.
Specifically, $\bm A$ is $(N_s\cdot L)\times M$ matrix, where $l\in[1,L]$, and $i\in[1,N_s]$,
\begin{equation}
\left(\bm A\right)_{N_s(l-1)\!+\! i,m}= \sum_{n\in\mathcal N_l} \! \frac{\lambda\cdot r_{n,m}(\hat{s}_i)\!\cdot\! \sqrt{g_{T,n}g_{R,m}}\!\cdot\! e^{-j2\pi (d_n \!+ \!d_{n,m})/\lambda}}{4\pi d_{n}d_{n,m}},\nonumber
\end{equation}
where 
$r_{n,m}(\hat{s}_i)$ denotes the reflection coefficient of the $n$-th {\ris} element for the incident signal towards the $m$-th space block in the $i$-th state,
$g_{T,n}$ is the gain of the transmitter towards the $n$-th {\ris} element,
$g_{R,m}$ is the gain of the receiver towards the $m$-th space block,
$d_n$ is the distance from the Tx to the $n$-th {\ris} element,
and $d_{n,m}$ denotes the distance from the $n$-th {\ris} element to the Rx antenna via the $m$-th space block.
Here, $r_{n,m}(\hat{s}_i)$ is calculated by using the CST microwave studio~\cite{Hirtenfelder2007Effective}.
Besides, $g_{T,n}$ and $g_{R,m}$ are obtained from the datasheets of Tx and Rx antennas, respectively.

% 解释一下r_{n,m}(\nu_l)是可以从CST通过仿真计算的。
% 首先将C转化为one-hot
We then transform $\bm C$ to a $K\times(L\cdot N_s) $ zero-one matrix $\bm D$, which satisfies $\forall k\in[1,K]$, $l\in[1,L]$, and $i\in[1,N_s]$,
\begin{equation}
\left( \bm D \right)_{k,N_s\cdot(l-1)+ i} = \begin{cases}
1,\quad \text{if $(\bm C)_{k,l}=n$},\\
0,\quad \text{otherwise}.\nonumber
\end{cases}
\end{equation}
Therefore, $\bm H$ can be calculated by $\bm H = \bm D \bm A$.
We denote process of calculating $\bm H$ from $\bm C$ by the mapping $\bm g: \bm C \rightarrow \bm H$.

 \end{appendices}

\vspace{2em}
\begin{small}
\bibliographystyle{IEEEtran}
\bibliography{ms.bib}
\end{small}

\end{document}